%% file: 000_main.tex
\documentclass[]{spie}  %>>> use for US letter paper
\input{packages}

\title{Investigations on assembly and coverage for modular focal planes of multiplexed telescopes}

\author[a]{Maxime Rombach}
\author[a]{Xiangyu Xu}
\author[b]{Ricardo Araujo}
\author[b]{Markus Thurneysen}
\author[c]{Stefane Caseiro}
\author[c]{Corentin Magnenat}
\author[d]{Joseph Silber}
\author[a]{Malak Galal}
\author[d]{David Schlegel}
\author[a]{Jean-Paul Kneib}
\affil[a]{Ecole Polytechnique Federale de Lausanne (EPFL), Route Cantonale, 1015 Lausanne, Switzerland}
\affil[b]{Haute Ecole du Paysage, d'Ingénierie et d'Architecture de Genève (HEPIA), Rue de la prairie 4, 1202 Geneva, Switzerland}
\affil[c]{Micro Precision Systems (MPS), Ch. du Longchamp 95, 2504 Bienne, Switzerland}
\affil[d]{Lawrence Berkeley National Laboratory (LBNL), 1 Cyclotron Road, Berkeley, CA 94720, USA}

\authorinfo{Further author information: (Send correspondence to M. R.)\\M.R.: E-mail: maxime.rombach@epfl.ch, Telephone: +41 (0) 21 693 28 76}

\begin{document} 
\maketitle
\thispagestyle{empty}

\begin{abstract}
Multiplexed surveys have the ambition to grow larger for the next generation of focal plane instruments. Future projects such as Spec-S5, MUST, and WST have an ever-growing need for multi-object spectroscopy (13,000 - 20,000 simultaneous objects) which demands further investigations of novel focal plane instrumentation. In this paper, we present a rigorous study of focal plane coverage optimization and assembly of triangular modules of alpha-beta fiber positioners with a 6.2 mm pitch.\\
The main focus here is to examine different module arrangements namely, framed, semi-frameless, and fully-frameless assemblies. Framed and semi-frameless describe here the usage of a manufactured focal plate to hold the modules together and provide the correct focus and tilt to the fibers. Work on automatically generating such focal plates for project adaptability and ease of manufacturing will also be presented. On the other hand, the frameless approach proposes a connection method freed from the need of a focal plate. The following paper will also present their capabilities to meet the requirements for focal plane assembly such as focus, tilt and coverage.
\end{abstract}


\input{001_introduction}

\input{002_FP_charac}
\input{003_Investigated_solutions}
\input{004_Evaluating_solutions}
\input{005_conclusion}

\acknowledgments % equivalent to \section*{ACKNOWLEDGMENTS}       
The authors acknowledge the support from the Innosuisse - Swiss Innovation Agency under Grant Agreement No. 101.014 IP-ENG. In addition, the authors would like to acknowledge Nicholas Wenner for the fruitful discussions and his insightful comments.
 
%Non exhaustive list yet and will become a paragraph later:
%\begin{itemize}
%    \item Along with the mentionned authors, Nicholas Wenner, and DESI team whose advice and work served as baseline for the presented investigations
%    \item \textit{Innosuisse} funding this project
%    \item EPFL ressources
%\end{itemize}

% References
\bibliography{SPIE_astrobots}
%\bibliography{references_spie} % bibliography data in report.bib
\bibliographystyle{spiebib} % makes bibtex use spiebib.bst

% I keep that here to have formatting procedures in at hand 
% Comment line below to remove formatting guidelines
% \include{100_formatting}

\end{document}

%% file: packages.tex
\usepackage{amsmath,amsfonts,amssymb}
\usepackage{booktabs}
\usepackage{float}
\usepackage{graphicx}
\usepackage{gensymb}
\usepackage[colorlinks=true, allcolors=blue]{hyperref}
\usepackage{subcaption}
\newcommand{\q}[1]{``#1''}
\usepackage{lmodern}
\usepackage{siunitx}

%% file: 001_introduction.tex
\section{INTRODUCTION}
\label{sec:intro}  % \label{} allows reference to this section

In a multi-fiber instrument, the Focal Plane System interfaces between the path of the light in free space and the fibers guiding it towards the spectrographs. Following the success of previous multi-fiber instruments such as SDSS-V \cite{sanchez-gallego_multi-object_2020} and DESI \cite{collaboration_overview_2022}, the next generation of instruments aims to observe an increasingly large number of celestial objects. Three main projects are currently being discussed within the community for a first light in a time frame of about 10 years:
\begin{itemize}
    \item Spec-S5: telescope and instrument project, follow-up of DESI \cite{collaboration_overview_2022}, mounted on two telescopes for both hemispheres and lead by the Lawrence Berkeley National Laboratory (LBNL)
    \item MUltiplexed Survey Telescope (MUST)\cite{zhang_conceptual_2023}: 6.5 m telescope and instrument project lead by Tsinghua University
    \item  Wide field Spectroscopic Telescope (WST) \cite{vernet_wst_2024}:  12 m telescope and instrument project, lead by European institutions 
\end{itemize}

\begin{table}[H]
\centering
\caption{Targets of future multi-fiber instruments}
\label{tab:objects_projects}
\begin{tabular}{lcccc}
\hline
                  & Spec-S5   & MegaMapper & MUST   & WST    \\ \hline
Primary mirror diam   & 2 $\times$ 6 m     & 6.5 m & 6.5 m   & 12 m    \\
\# objects/robots & 2 $\times$ 13,000 & 20,000 & 21,000 & 20,000 \\ \hline
\end{tabular}
\end{table}

This diversity of ambitious projects highlights the need for new solutions in their development. We will study here the investigations realized for the assembly of their fiber positioners. The work baseline starts from the \textit{MegaMapper (MM) concept} \cite{blanc_megamapper_2022}, that proposed a \textit{modular assembly} of fiber positioners. As opposed to previous projects, the robots will not be assembled one by one, but will be assembled by full modules or rafts containing several of them; typically 63 robots per module as seen later in Section \ref{sec:nb_robots} Each robot is separated by 6.2 mm from its neighbors' (\textit{pitch} distance) and carries one optical fiber.

\begin{figure}[H]
\begin{subfigure}[b]{0.5\textwidth}
		\centering
		\includegraphics[width=0.5\linewidth]{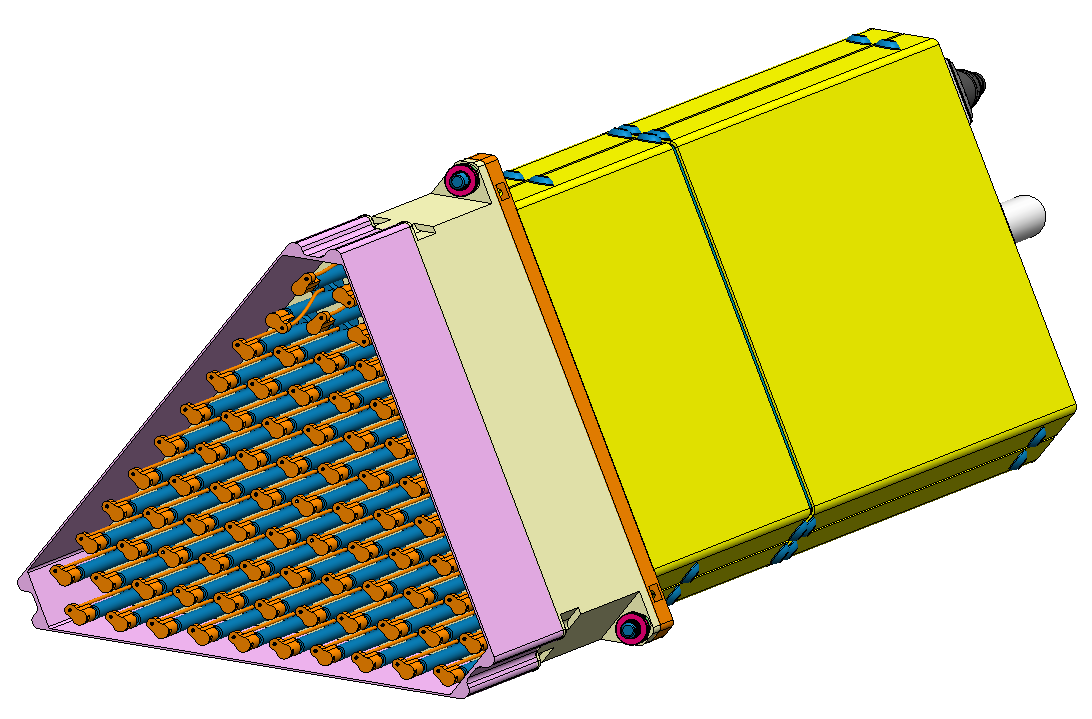}
		\caption{MM: Base module design with 75 SCARA robots}
		\label{fig:MM_module_concept}
	\end{subfigure}
	\begin{subfigure}[b]{0.5\textwidth}
		\centering
		\includegraphics[width=0.7\linewidth]{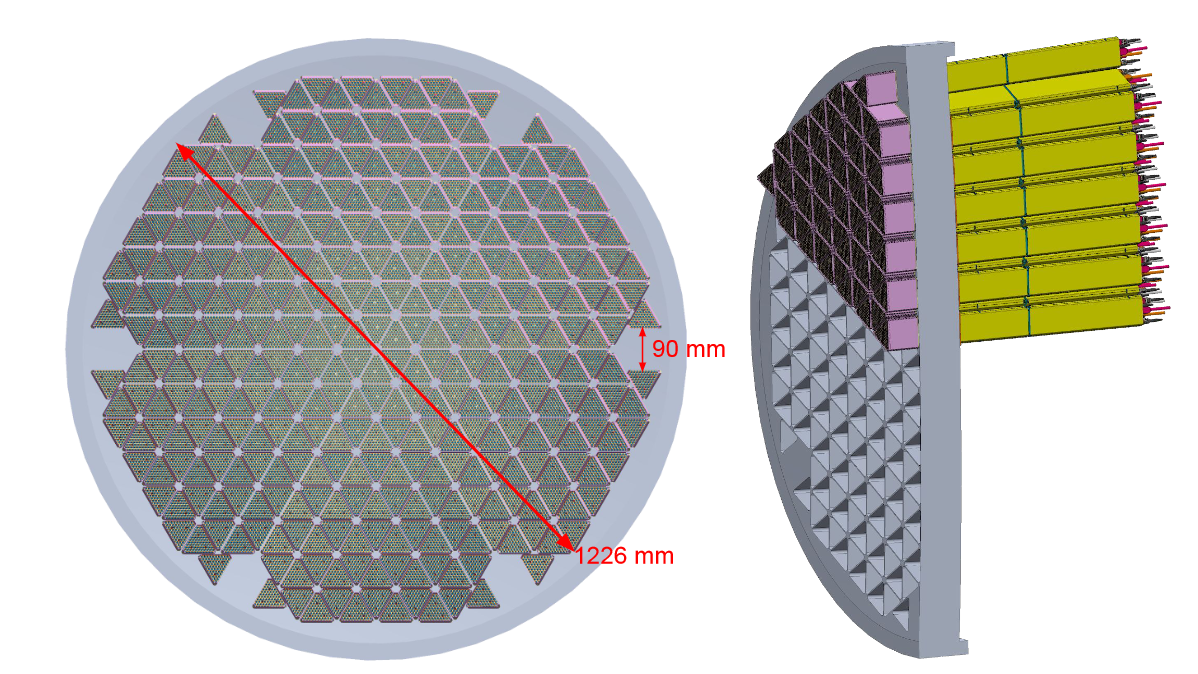}
		\caption{MM: focal plane assembly\cite{silber_25000_2022}}
		\label{fig:MM_FP_assembly}
	\end{subfigure}
  \vspace{0.3cm}
	\caption{MegaMapper focal plane concept}
	\label{fig:MM_concept}
\end{figure}

The following sections have two objectives: 1) To give a general overview of relevant design parameters for a telescope focal plate, 2) To propose a guidelines on how to advance the MM concept further by firslty optimizing positioners' coverage, and secondly by come up with different concepts for the module assemblies. The strength of a modular assembly lies in its adaptability to different projects. While the results in Section \ref{sec:evaluating_sol} use the focal surface parameters of the latest Spec-S5 update, this modular solution can be adapted to any highly multiplexed fibers project. Hence, similar studies can easily be conducted for the MUST or WST projects for example.

%Begin the Introduction below the Keywords. The manuscript should not have headers, footers, or page numbers. It should be in a one-column format. References are often noted in the text and cited at the end of the paper.

%% file: 002_FP_charac.tex
\section{Focal plane characteristics impacting design}
\label{sec:FP_charac}
Before delving deeper into presenting the design itself, it is important to thoroughly understand the core characteristics of a focal plane impacting its design and optimization. The trade-off considered here will be the balance found between the maximized coverage of the positioners and the tilt/focus parameters.

\subsection{Inputs}
To reach an optimized design, there a few vital parameters that can be tuned and adjusted according to the specifications of each project. The parameters are the following: the focal surface of the project, the number of robots per module, and the module gaps.
\subsubsection{Focal surface parameters}
As already mentioned earlier in Section \ref{sec:intro}, one of the main strengths of a modular assembly is its multi-project aspect. The focal surface parameters define the (a)spherical surface that the modular assembly should match. They are project dependent and influence for instance the maximum allowable number of modules.\\
The studies presented in this paper are based on the most recent parameters from the Spec-S5 project which has a \textit{spherical} focal surface.

\begin{table}[H]
\centering
\caption{Focal surface parameters for Spec-S5 and MUST}
\label{tab:spec-s5_param}
\begin{tabular}{@{}ccccc@{}}
\toprule
        & Best fit sphere {(}mm{)} & Vignetting radius {(}mm{)} & f-number (-) & Surface shape \\ \midrule
Spec-S5 &   12657                      &  409.4                          &  3.62002        & spherical     \\
MUST    & 10477.594                & 592.35                     & 3.72059  & aspherical    \\ \bottomrule
\end{tabular}
\end{table}
\subsubsection{Number of robots per module}
\label{sec:nb_robots}
\begin{figure}[H]
\captionsetup[subfigure]{justification=centering}
\begin{subfigure}[b]{0.33\textwidth}
		\centering
		\includegraphics[width=0.72\linewidth]{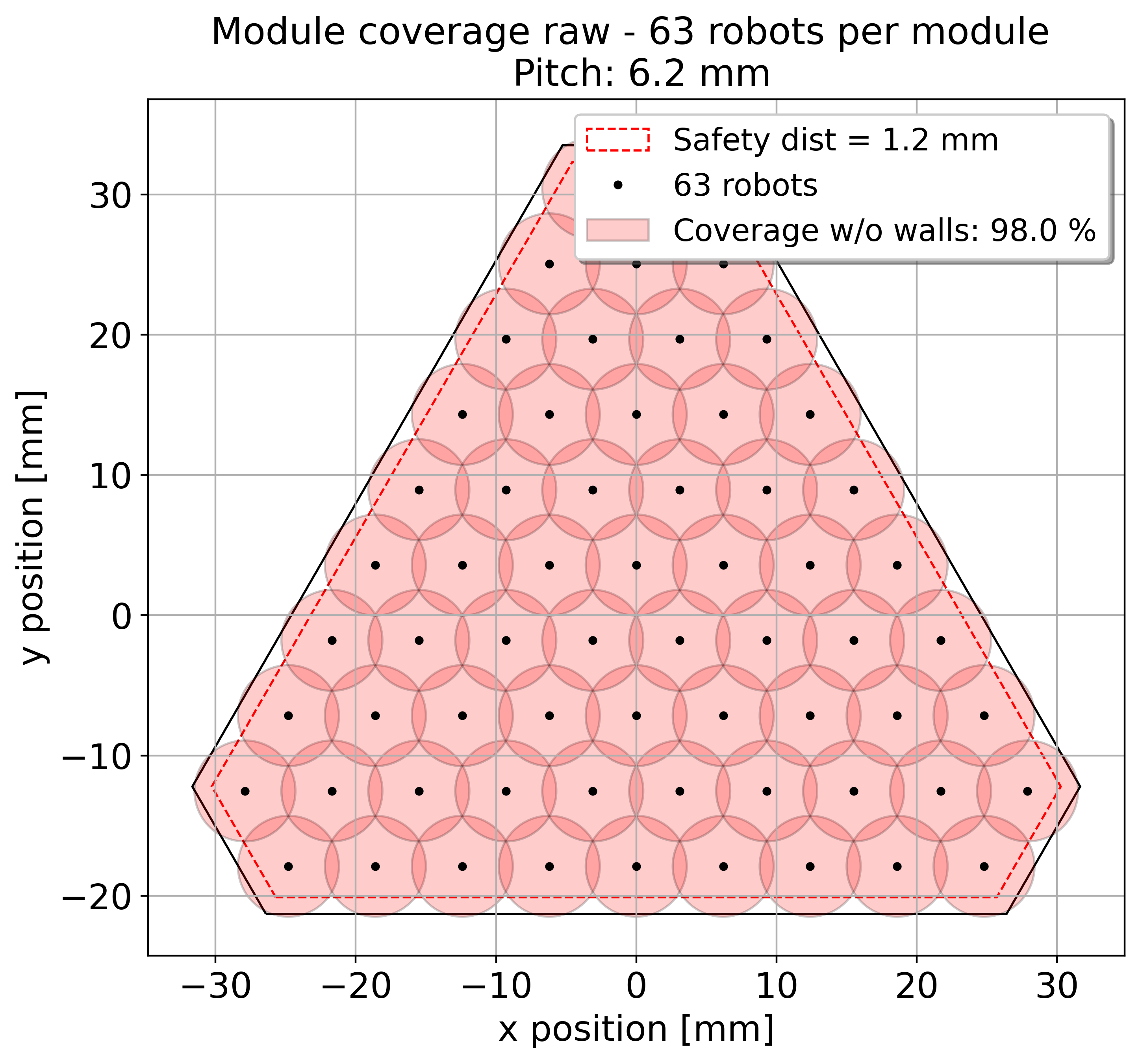}
		\caption{63 robots with individual workspaces\\ Module side length: 73.8 mm}
		\label{fig:63_robots}
	\end{subfigure}
	\begin{subfigure}[b]{0.33\textwidth}
		\centering
		\includegraphics[width=0.72\linewidth]{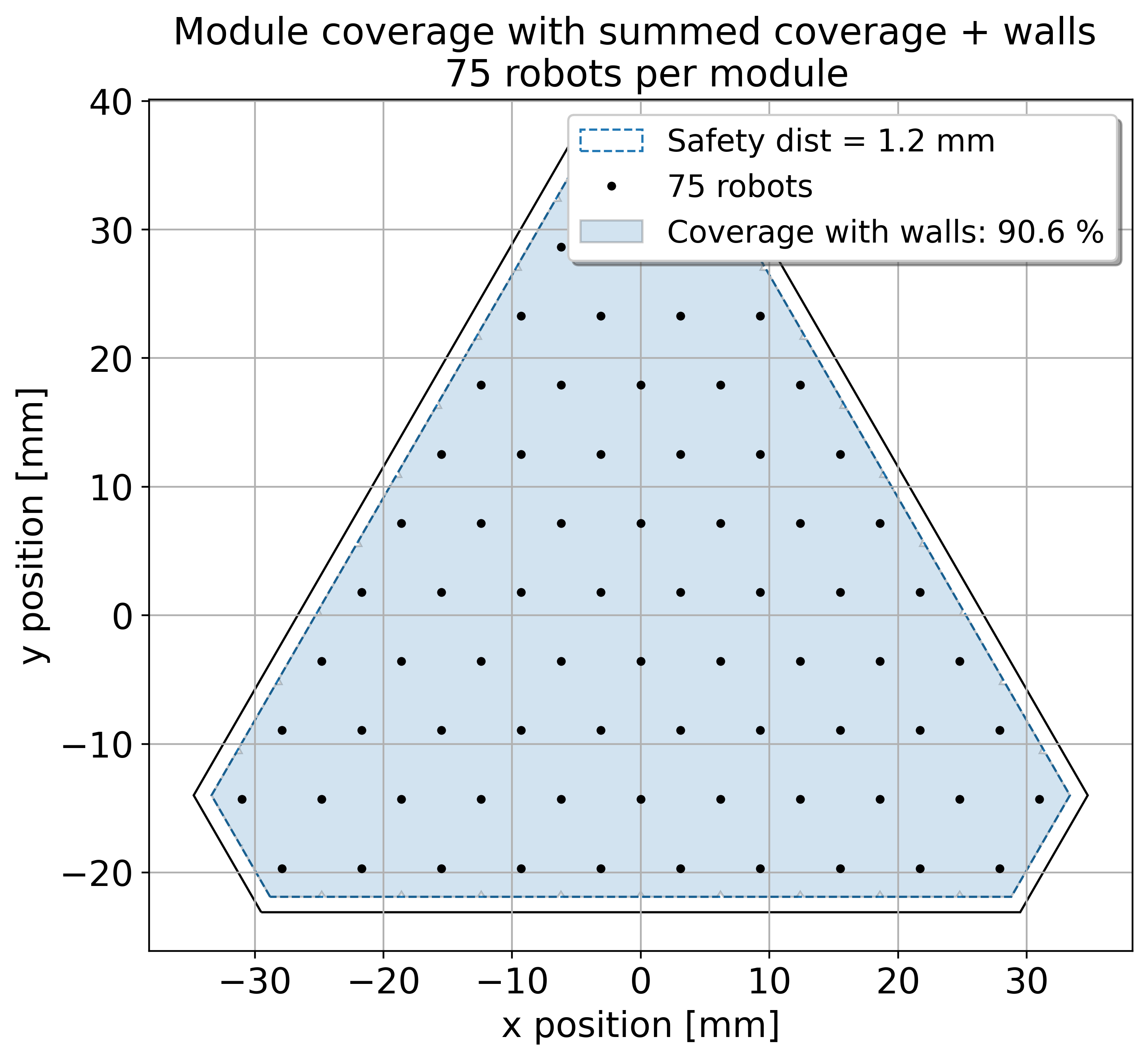}
		\caption{75 robots with summed up workspaces\\ Module side length: 80 mm}
		\label{fig:75_robots}
	\end{subfigure}
 \begin{subfigure}[b]{0.33\textwidth}
		\centering
        \includegraphics[width=0.72\linewidth]{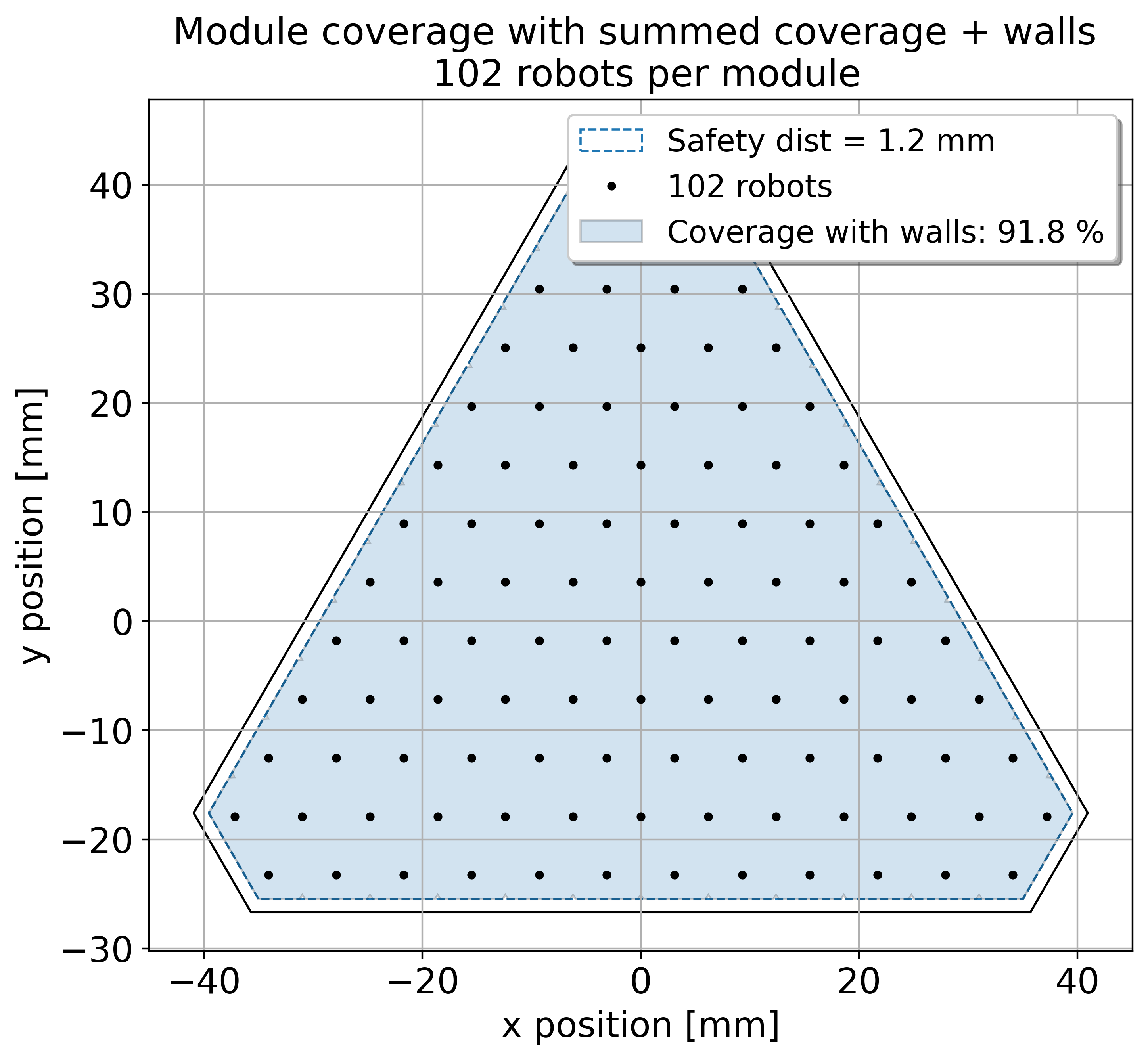}
		\caption{102 robots with summed up workspaces\\ Module side length: 92.4 mm}
		\label{fig:102_robots}
	\end{subfigure}
    \vspace{0.3cm}
	\caption{Varying number of robots inside a module and their coverage; each robots has a patrol radius of 3.6 mm and a pitch of 6.2 mm}
	\label{fig:number_of_robots_per_module}
\end{figure}
The number of robots per module influences the final coverage of the focal surface. Larger modules provide a better paving but at the cost of loss in positioning precision of the optical fibers as seen later in Section \ref{sec:evaluation_number of robots per module}.\\
The number changes by adding or removing rows of positioners. The <<75>> case was the baseline of the MegaMapper concept. We switch from one case to another by adding or removing rows of positioners. One row is removed to obtain the <<63>> case, two rows added for the <<102>> case. Decision on this number is multi-factor. It depends on the coverage versus focal surface fitting trade-off, but also on how easy each module will be to assembled by the manufacturer or if it can fit an integer number of \textit{trilliums}, units of three robots, developed in LBNL \cite{blanc_megamapper_2022}. The latter is one option for fitting positioners in the modules, and other investigations are conducted to still assemble individual robots instead of units of three. For fitting an integer number of trillums, one module needs house a multiple of 9 robots. The 63 case fulfills this requirement but not the 75 and 102 ones. Adding three more rows to reach 117 robots fulfills it as well but, as it will be shown in Section \ref{sec:evaluation_number of robots per module}, increasing too much the number of robots results in a worse fitting of the focal surface. Consequently, the highest number of robots to be considered is 102 robots.

\begin{figure}[H]
    \centering
    \includegraphics[width=0.35\linewidth]{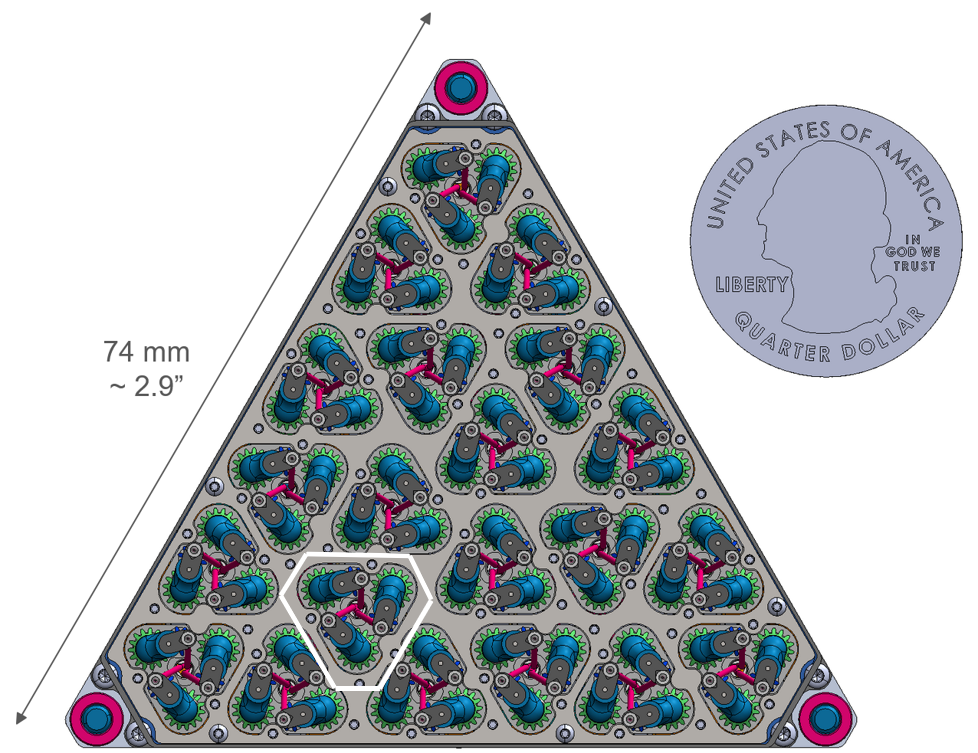}
    \vspace{0.1cm}
    \caption{21 trillium pattern (63 positioners) wiht one unit highlighted in white. Credit: Joseph Silber and Nicholas Wenner (LBNL)}
    \label{fig:trillium_pattern}
\end{figure}

\subsubsection{Module gaps and out allowance}
\begin{figure}[H]
\captionsetup[subfigure]{justification=centering}
\begin{subfigure}[b]{0.5\textwidth}
		\centering
		\includegraphics[width=0.8\linewidth]{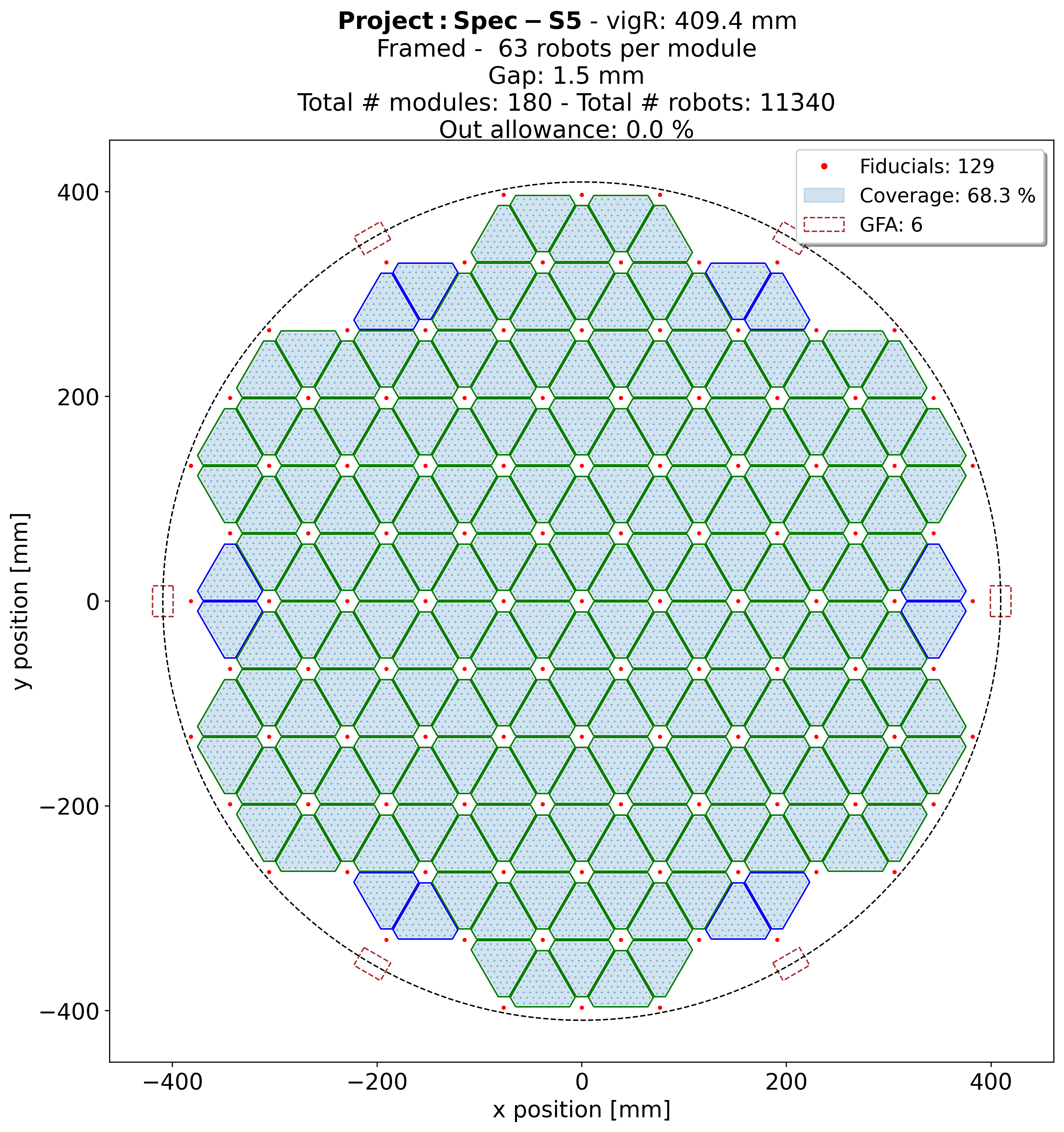}
		\caption{Paving with 1.5 mm gap between modules \\ and 0\% of out allowance}
	\end{subfigure}
	\begin{subfigure}[b]{0.5\textwidth}
		\centering
		\includegraphics[width=0.8\linewidth]{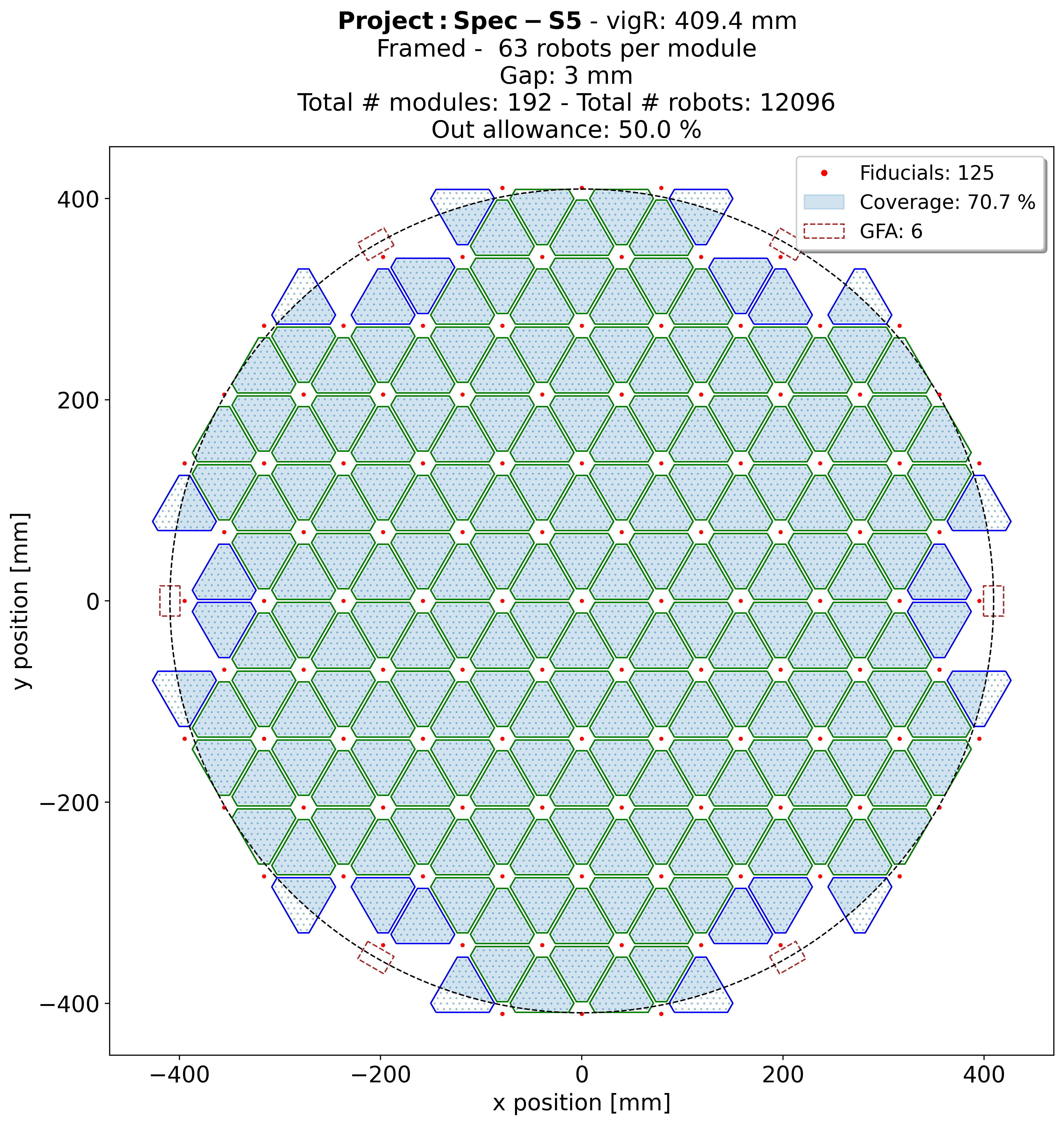}
		\caption{Paving with 3 mm gap between modules \\ and 50\% of out allowance}
	\end{subfigure}
  \hspace{0.3cm}
	\caption{Two cases of focal plane filling to highlight the impact of module gap and out allowance}
	\label{fig:module_gap_out_allow}
\end{figure}

The module gaps along with the out allowance are the two fundamental parameters that directly impact the modules arrangement on the focal surface. The first parameter defines how the sides of two neighboring modules are far apart from each other. \textit{Global} and \textit{Inner gaps} are going to be explained in more detail in Section \ref{sec:semi-frameles solution}\\
While one could limit the paving of the focal surface to remain strictly within the \textit{vignetting radius}, i.e. the region where the light quality is optimum, this approach limits greatly the coverage potential of the paving. Indeed, if one allows for a portion of the modules to stick out of the vignetting radius, more robots can be included on the focal surface. The maximum percentage of module area allowed to stick out of the focal vignetting radius is called the \textit{out allowance}. Its impact on the focal plane layout is illustrated in Figure \ref{fig:module_gap_out_allow}. The different layout that are presented in this study are similarly calculated for DESI \cite{silber_robotic_2022} and MM\cite{silber_25000_2022}.

\subsubsection{Guide, Focus and Alignment cameras and wavefront sensors}
\label{sec:pres_GFAs}
\begin{figure}[H]
\captionsetup[subfigure]{justification=centering}
\begin{subfigure}[b]{0.33\textwidth}
		\centering
		\includegraphics[width=\linewidth]{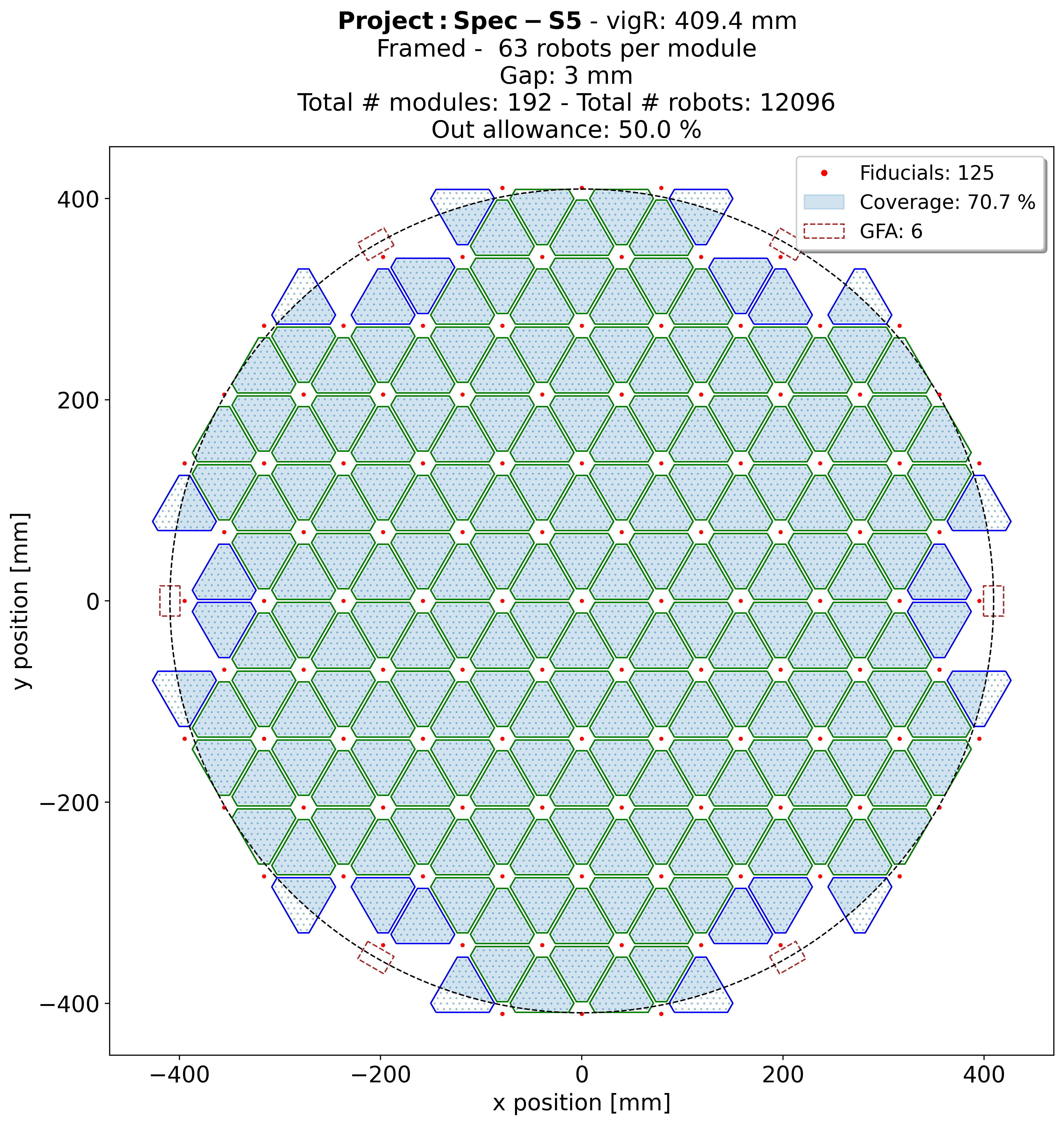}
		\caption{6 minimized GFAs footprints: 30x20 mm - Coverage: 70.7\%}
	\end{subfigure}
	\begin{subfigure}[b]{0.33\textwidth}
		\centering
		\includegraphics[width=\linewidth]{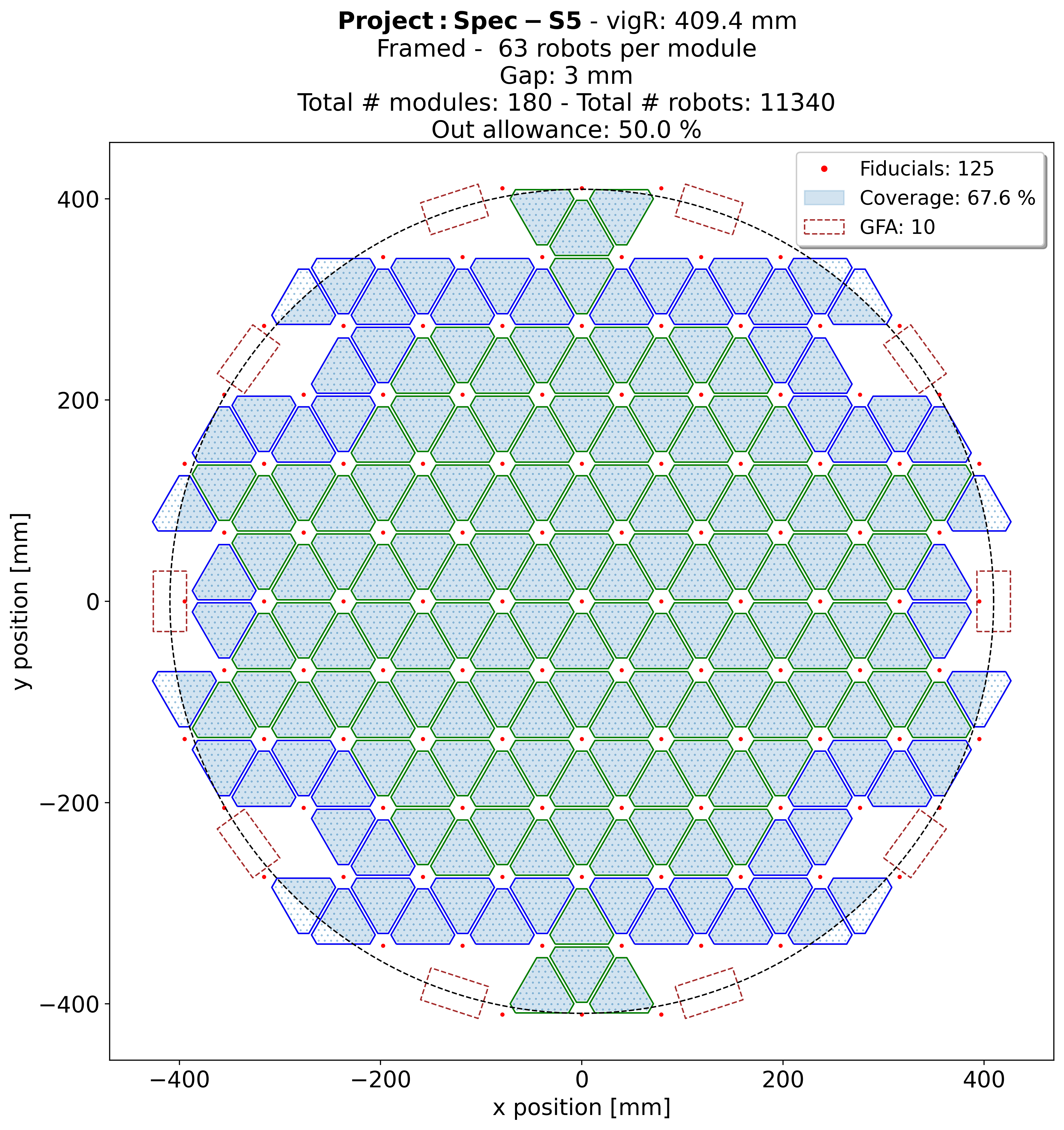}
		\caption{10 DESI-like GFAs footprints: 61x33.3 mm - Coverage: 67.8\%}
	\end{subfigure}
 \begin{subfigure}[b]{0.33\textwidth}
		\centering
		\includegraphics[width=\linewidth]{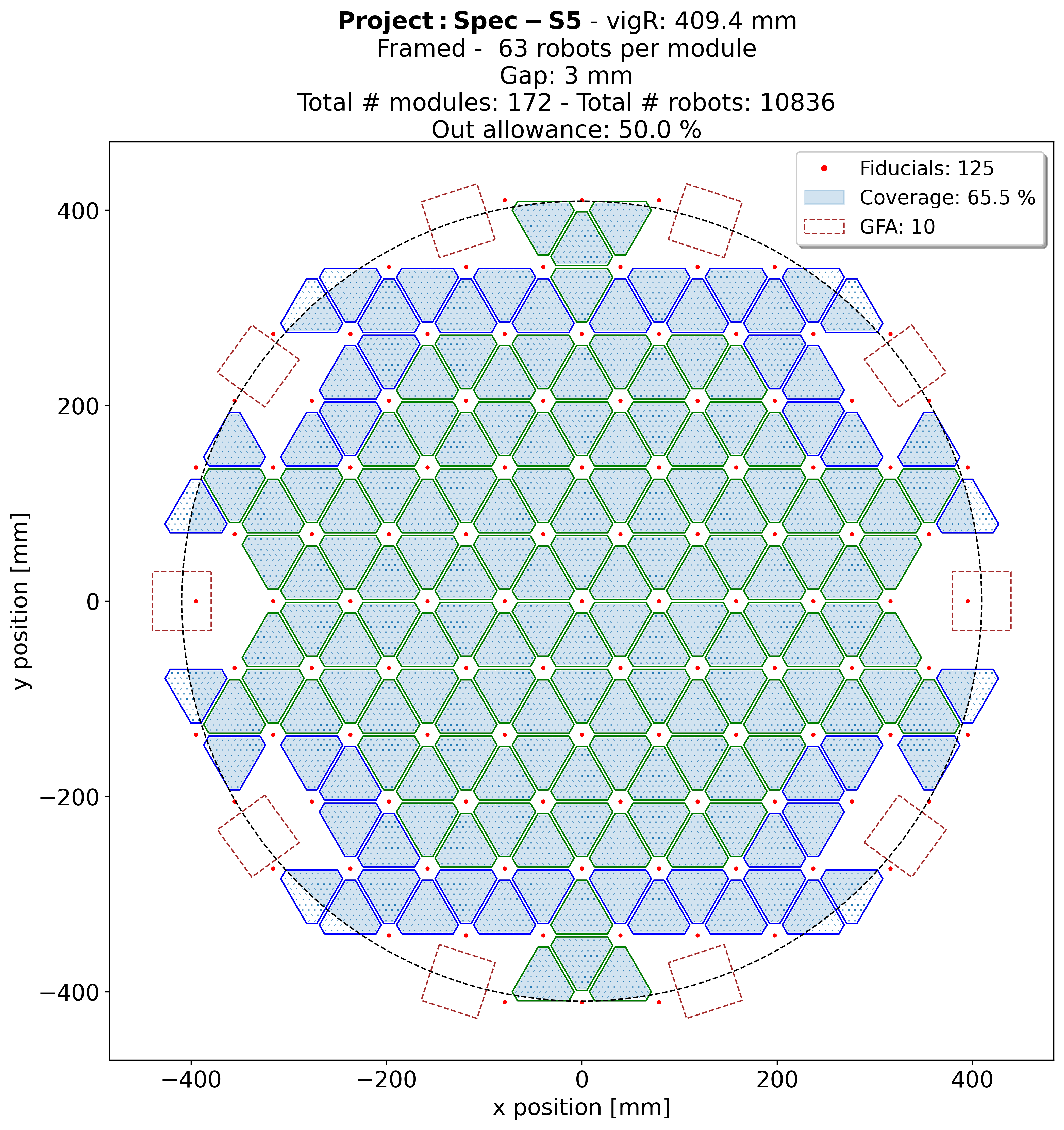}
		\caption{10 large GFAs footprints: 60x60 mm - Coverage: 65.5\%}
	\end{subfigure}
  \hspace{1cm}
	\caption{Framed layout (global gap: 3 mm) comparison for different sizes and number of GFAs to highlight their impact}
	\label{fig:GFAs_comparison}
\end{figure}

Encapsulated later in the term \textit{GFAs}, those sensors are necessary for observations and have a significant impact on the focal plane layout depending on their footprint, numbers and position. Contrary to the vignetting radius they correspond to a physical limit that a module cannot overlap. If this occurs in the layout, the overlapping modules are removed, losing as many robots at once. Figure \ref{fig:GFAs_comparison} illustrates that problematic.To date, those are the most unknown parameters impacting the modules layout. Each future project has not yet defined their design. Assumptions have, then, to be made on their number, size and position. In order to assess the maximum potential of a modular layout, decision was made to minimize their number and size while still taking them into account. Therefore, six 30$\times$20 mm GFAs footprints are placed on the vignetting circle similarly to SDSS-V for instance.

\subsection{Intrinsic properties: defocus and tilt}
\begin{figure}[H]
\captionsetup[subfigure]{justification=centering}
\begin{subfigure}[b]{0.5\textwidth}
		\centering
		\includegraphics[width=0.7\linewidth]{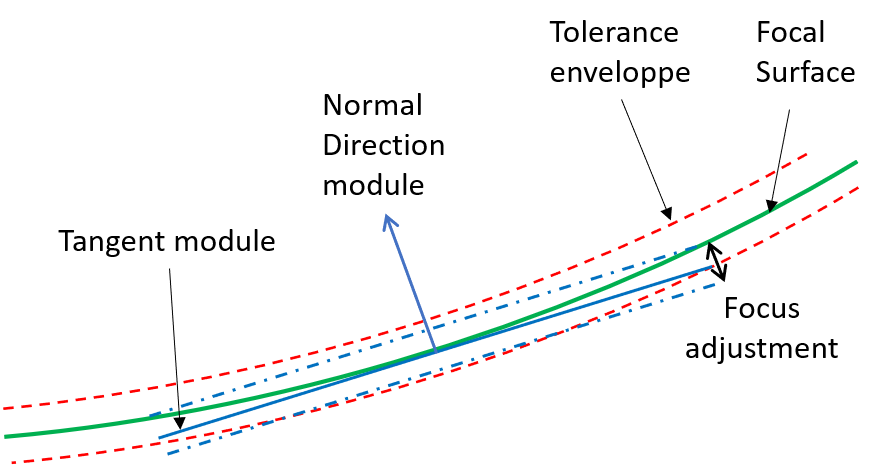}
		\caption{Defocus tolerance}
	\end{subfigure}
	\begin{subfigure}[b]{0.5\textwidth}
		\centering
		\includegraphics[width=0.7\linewidth]{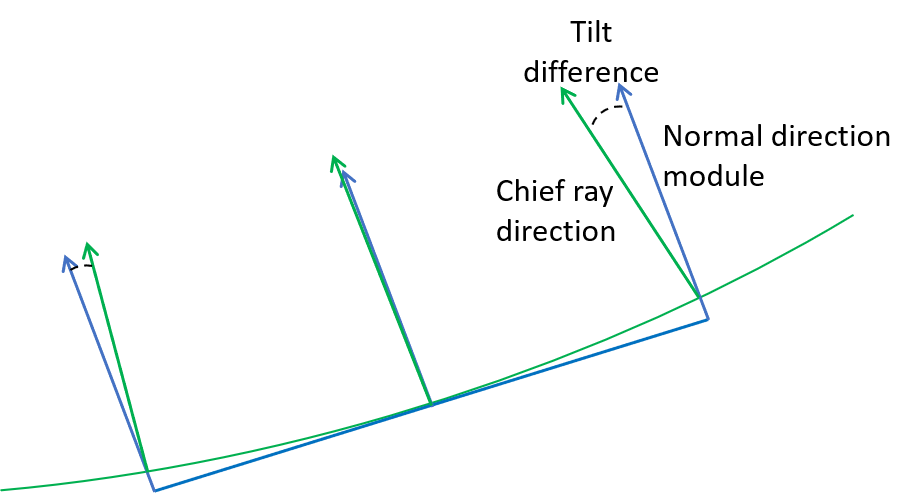}
		\caption{Tilt}
  \label{fig:tilt_schematics}
	\end{subfigure}
  \hspace{0.3cm}
	\caption{Focus and tilt tolerances illustration}
	\label{fig:focus&tilt}
\end{figure}

% Intrinsic characteristics of the focal surface/ telescope that define the nominal values of the required positioning/ orientation accuracy of the fibers (focus and tilt adjustment).\\
On the other side of the considered trade-off with coverage lies the focus and tilt adjustment of the optical fiber tips. Those two parameters are given by the optical design of the telescope and define the nominal targets that the fiber tips should reach. They are critical as they set \textit{how well} and \textit{how much} light enters the optical fibers, thus directly impacting the quality of the signal received at the end of the fiber route by the spectrographs. Figure \ref{fig:focus&tilt} highlights those two properties assuming that all the fiber tips lie on a plane at the edge of the module called later as \textit{fiber tips plane}.\\

\subsection{Output}
\subsubsection{Coverage}
The coverage is the ratio of the area patrolled by the fiber positioners inside the focal surface over the total available area depending on the instrument project. The target coverage to aim for is then defined by the science requirements of said projects.\\
It is important to keep in mind that in past projects each robot was assembled individually on the focal plane and thus easing the coverage process. The modular solution \textemdash despite it not exactly fulfilling the simplicity of the coverage process \textemdash overweighs the individual assembly by offering two significant advantages:
\begin{itemize}
    \item a remarkably higher robot density with a target to 20,000 robots, 4$\times$ more than the highest density to date
    \item a considerably lower instrument downtime for maintenance as the modularity allows for faster robot replacement
\end{itemize}
\subsubsection{Deflection/ Tilt}
The optical fiber tips at the end of each module need to be in focus and aligned with their corresponding chief ray of light. Therefore, the module layout design needs to result in a stiff enough assembly that will deform a little as possible to maintain the fiber tips in place. Finite Element Analyses (FEAs) are performed to assess such deflection and tilts in Section \ref{sec:fea} for various focal plate designs.
\subsubsection{Ease of assembly}
This last qualitative parameter evaluates how easy the assembly and disassembly processes are. It will turn into a quantitative one once the future projects start assembly tests and time assessment.

%% file: 003_Investigated_solutions.tex
\section{Investigated solutions}
\label{sec:investigated_solutions}
Based on the presented parameters in Section \ref{sec:FP_charac} it is now possible to present the 3 main envisioned assembly solutions for future highly multiplexed focal plane system.\\

\subsection{Frameless concept}
\begin{figure}[H]
\begin{subfigure}[b]{\textwidth}
		\centering
		\includegraphics[width=0.7\linewidth]{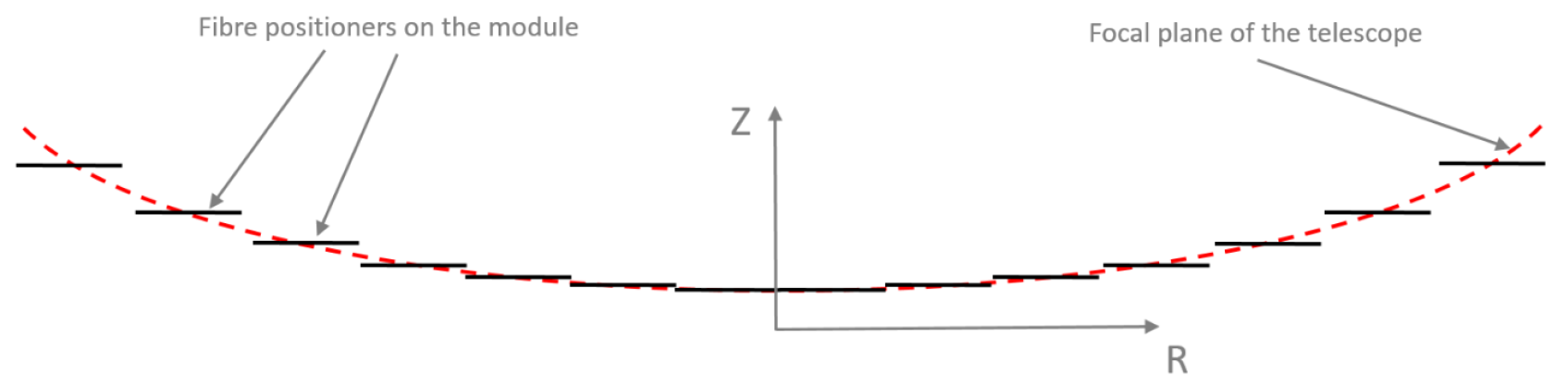}
		\caption{Staired assembly idea}
		\label{fig:staired_assembly}
	\end{subfigure}
\begin{subfigure}[b]{0.5\textwidth}
		\centering
		\includegraphics[width=0.5\linewidth]{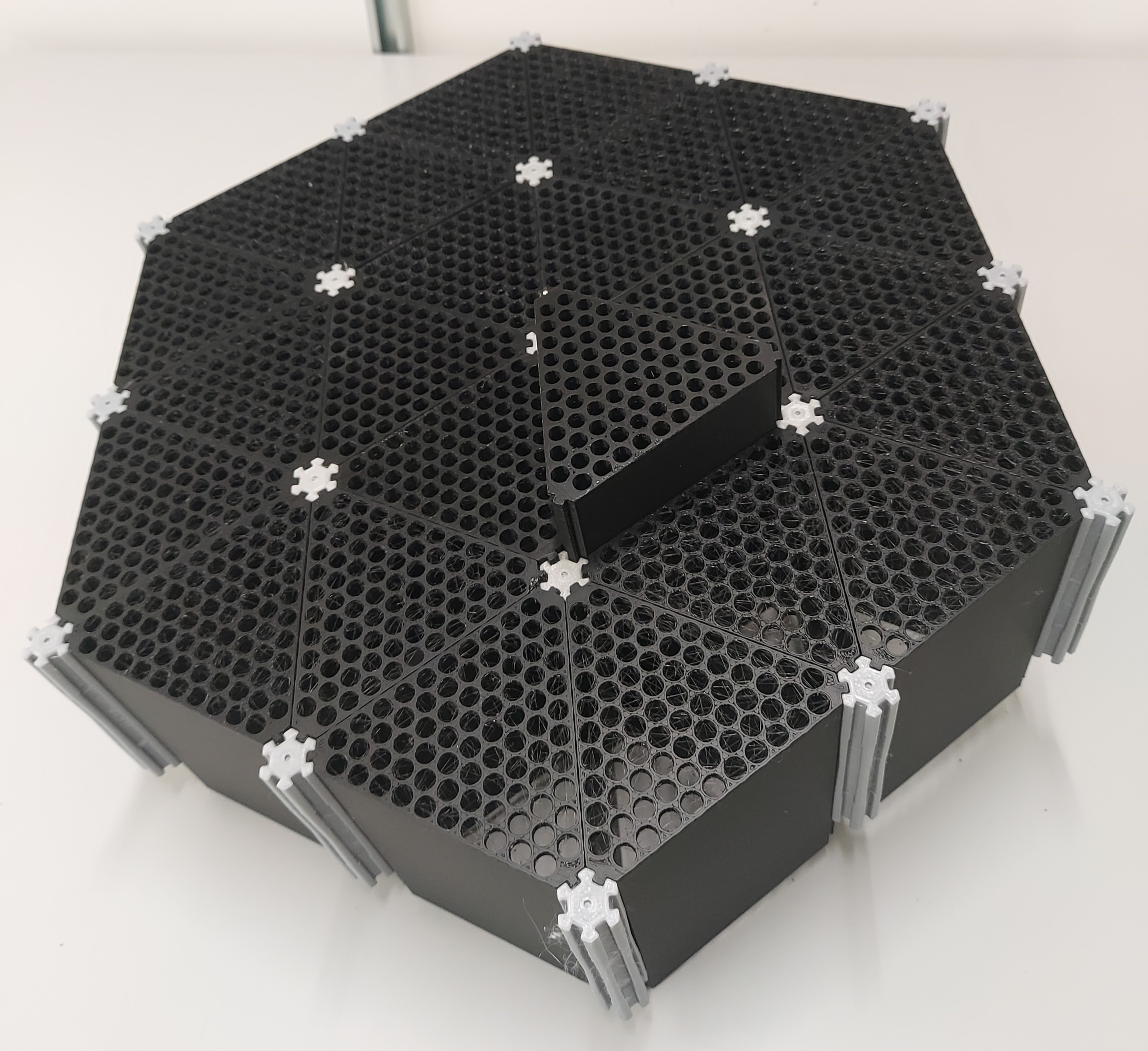}
		\caption{Frameless assembly concept}
		\label{fig:frameless_assembly}
	\end{subfigure}
	\begin{subfigure}[b]{0.5\textwidth}
		\centering
		\includegraphics[width=0.5\linewidth]{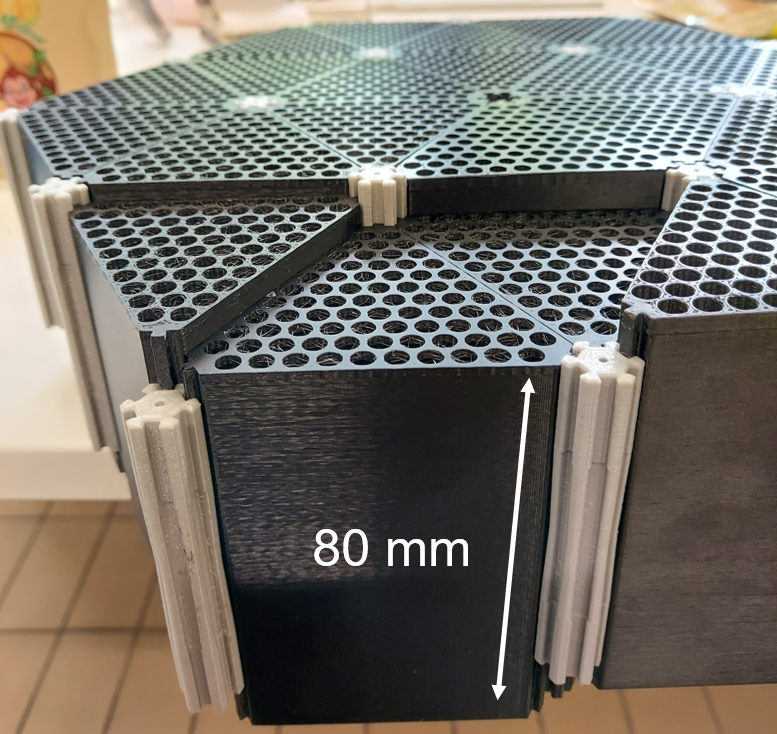}
		\caption{Close-up view on connection parts}
		\label{fig:frameless_connection part}
	\end{subfigure}
    \vspace{0.15cm}
	\caption{Frameless concept with 75 robots/module}
	\label{fig:frameless_concept}
\end{figure}

The current project was initiated to investigate the possibility of a \textit{frameless} assembly. The idea was to study the concept of an assembly that would no longer require the manufacturing of a supporting frame, hence a self-supporting structure. The thought of a \textit{staired} assembly of vertical modules was also to be investigated as proposed by the Innosuisse project proposal\cite{araujo_modular_2022} and is described in Figure \ref{fig:staired_assembly}. It will be shown later in Section \ref{sec:evaluating_sol} that a simple staired assembly of modules does not provide the necessary tilt to fit the focal surface.
Figure \ref{fig:frameless_concept} shows the 3D printed concept of modules interconnected tiny connection parts. The precision manufacturing needed is reduced from a large plate to small connection parts, reducing manufacturing risks and increasing the number of possible suppliers. Moreover, its modularity provides the freedom of assembling as many modules as one would require for a given project. Finally, removing the web between modules allows increase the coverage of the robots inside the focal plane as shown in Figure  \ref{fig:coverage_framed_frameless}. \\
The main downside found to this solution lies as well in the interconnection of the modules. As they are not referenced to a fixed frame anymore but to one another, the tolerance chain in positioning, including defocus and tilt, adds up from the first module placed to the last. As it will be seen in Section \ref{sec:evaluation_number of robots per module}, focal surface fitting requires a precise tolerancing in modules positions.
\subsection{Semi-frameless solution}
\label{sec:semi-frameles solution}
\begin{figure}[H]
\captionsetup[subfigure]{justification=centering}
\begin{subfigure}[t]{0.5\textwidth}
		\centering
		\includegraphics[width=0.6\linewidth]{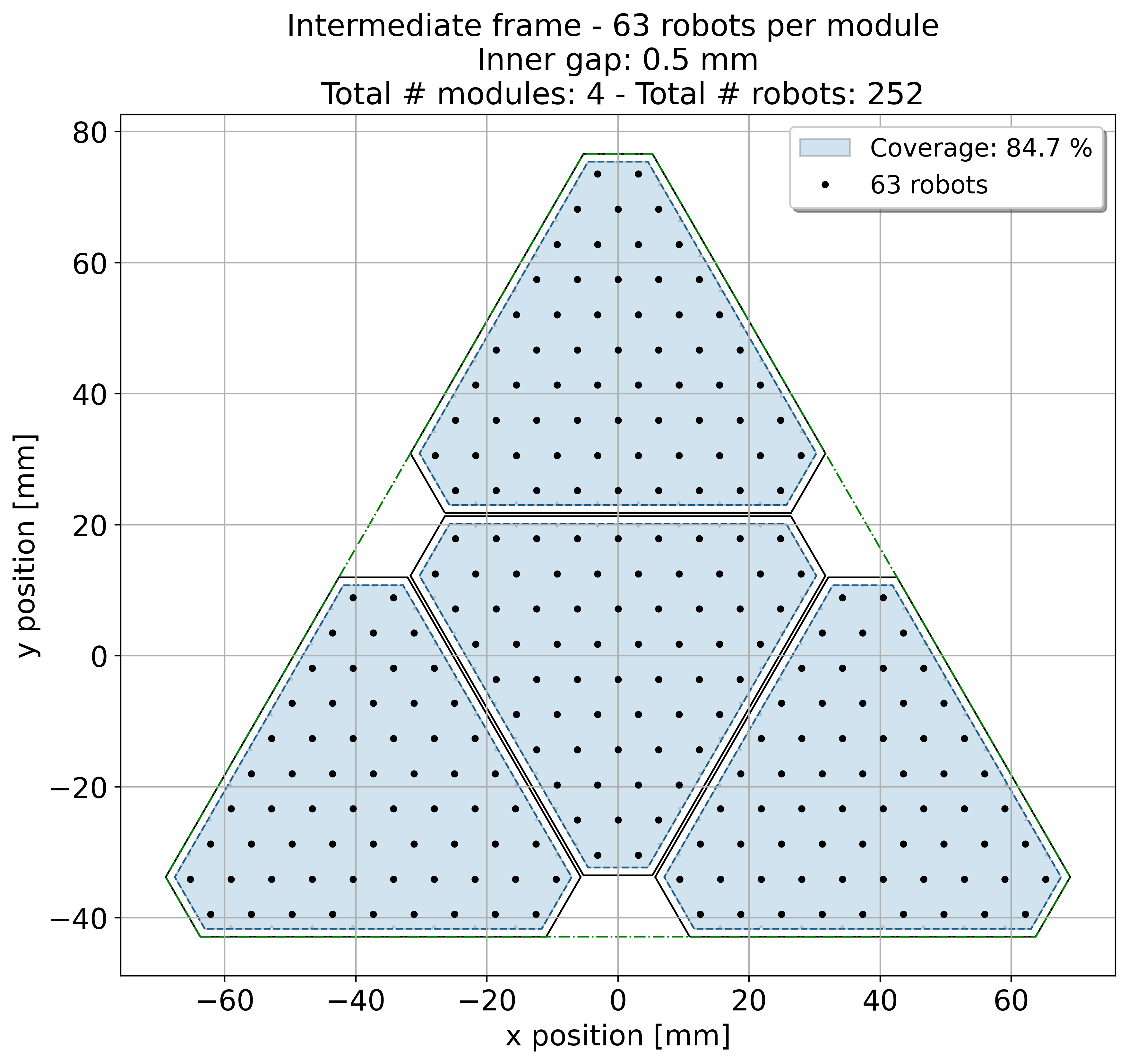}
		\caption{Intermediate triangle with locally closer modules of 63 robots\\}
	\end{subfigure}
	\begin{subfigure}[t]{0.5\textwidth}
		\centering
		\includegraphics[width=0.6\linewidth]{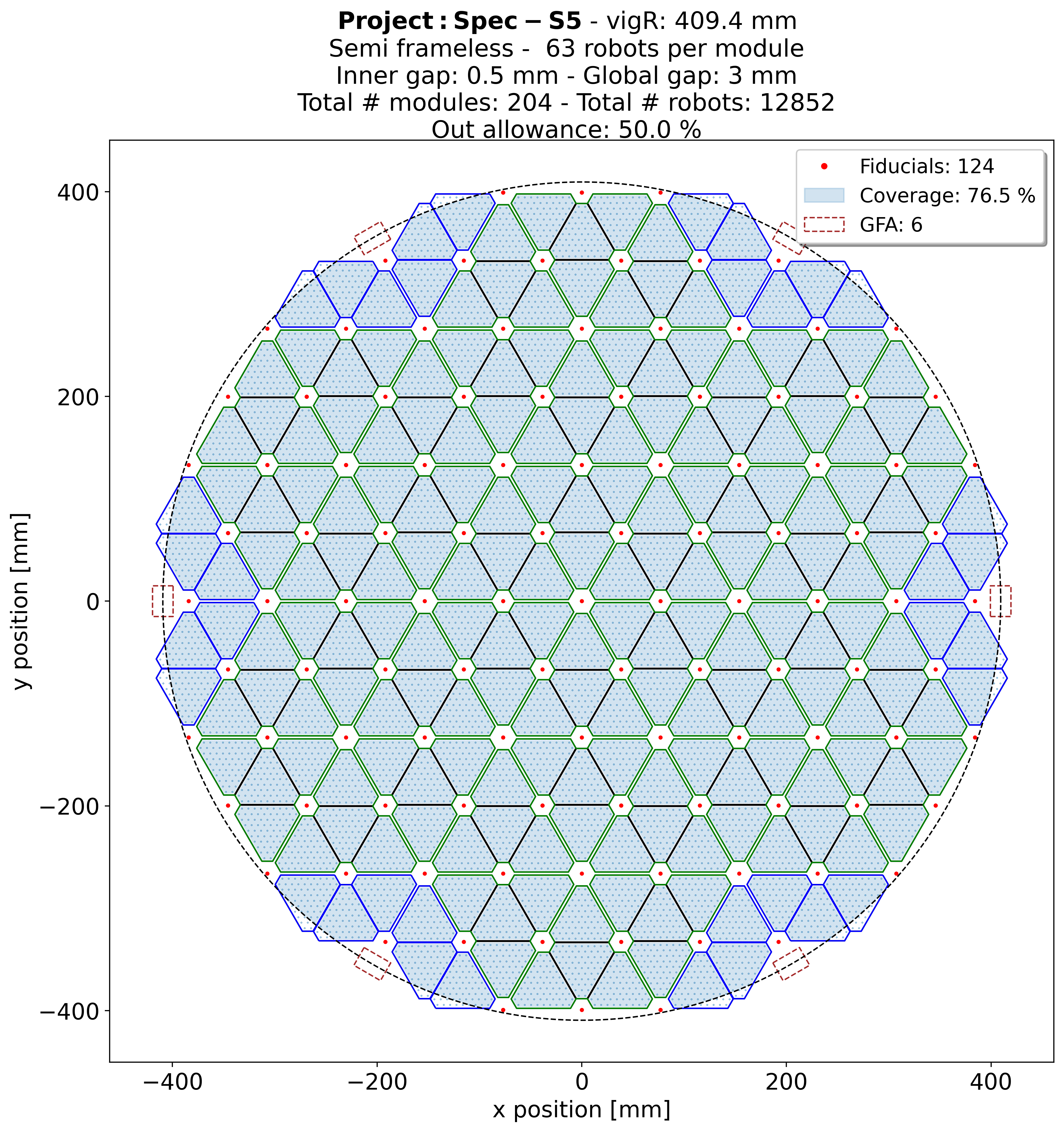}
		\caption{Semi-frameless assembly; green boundaries show full intermediate triangles (a) inside the vignetting, blue boundaries are partial intermediate triangles that fill the edge of the vignetting depending on the out allowance parameter\\
  Global gap: 3 mm - Inner gap: 0.5 mm}
	\end{subfigure}
  \vspace{0.3cm}
	\caption{Semi-frameless arrangement}
	\label{fig:inner_gap}
\end{figure}

Considering the frameless solution as the ideal case for coverage but not at all for robots positioning, one would propose an in-between solution: the \textit{semi-frameless}.\\
It proposes to keep the manufactured frame as a fixed positioning reference for all the modules but \textit{locally} bring some of them closer to increase the coverage. Walls inside a triangle of four modules are removed to allow for them to be closer. This new smaller gap is called \textit{inner gap} and is illustrated in Figure \ref{fig:inner_gap}.

\subsection{Framed solution}
\begin{figure}[H]
\captionsetup[subfigure]{justification=centering}
\begin{subfigure}[t]{0.5\textwidth}
		\centering
		\includegraphics[width=0.5\linewidth]{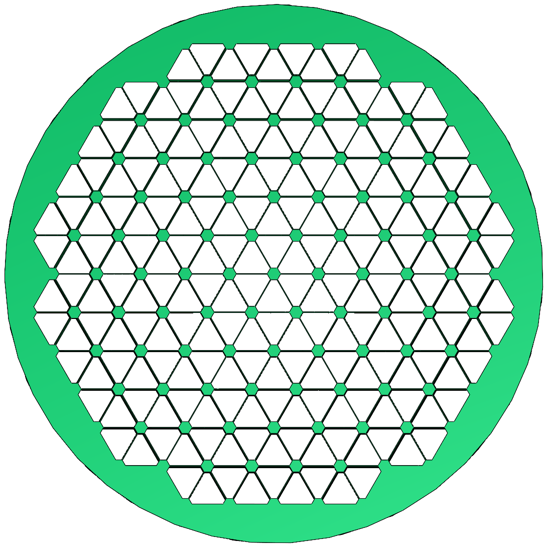}
		\caption{Top view of a \textit{framed} arrangement; full web with 1.5 mm walls}
	\end{subfigure}
	\begin{subfigure}[t]{0.5\textwidth}
		\centering
		\includegraphics[width=0.5\linewidth]{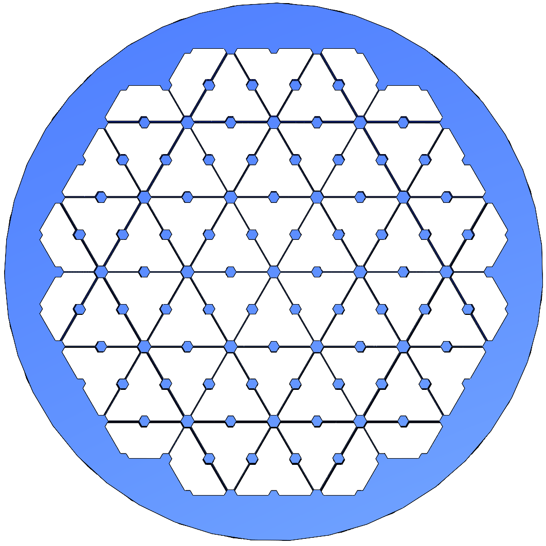}
		\caption{Top view of a \textit{semi-frameless} arrangement; partial web with 1.5 mm walls}
	\end{subfigure}
  \vspace{0.3cm}
	\caption{Visual comparison of the two main concepts; both have an outer diameter of 960 mm}
	\label{fig:framed_semi_frameless_cad}
\end{figure}

The basis of the current work and current frame solution is simply the full \q{web} of walls proposed in the MM concept described in Figure \ref{fig:MM_FP_assembly}. A visual comparison for clarity with the semi-frameless case is presented in Figure \ref{fig:framed_semi_frameless_cad}

%% file: 004_Evaluating_solutions.tex
\section{Evaluating solutions}
Now that the parameters, trade-offs and concepts have been established, it is possible to quantify their performances and draw conclusions.
\label{sec:evaluating_sol}
\subsection{Number of robots per module}
\label{sec:evaluation_number of robots per module}
\begin{figure}[H]
\captionsetup[subfigure]{justification=centering}
\begin{subfigure}[t]{\linewidth}
		\centering
		\includegraphics[width=0.8\linewidth]{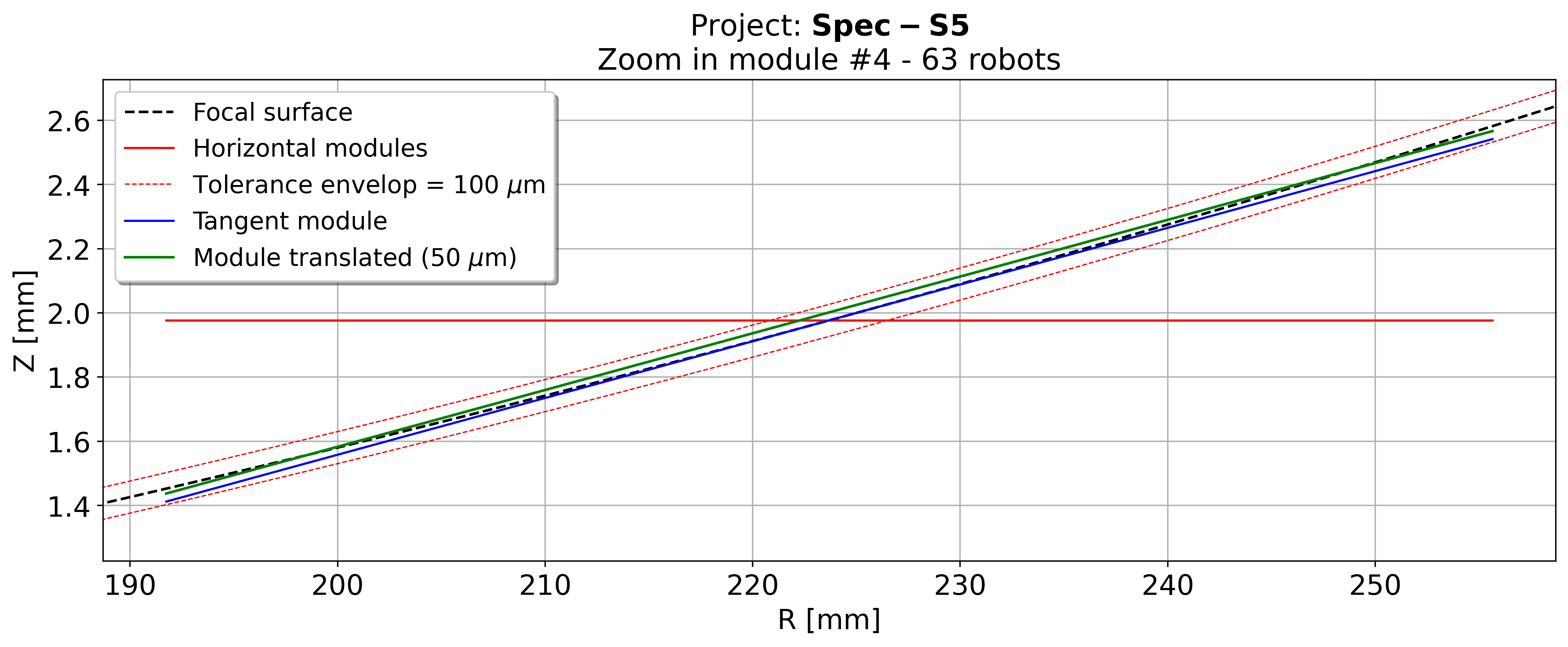}
		\caption{Positioning and orientation of one module at 222 mm from the center of the focal surface; plain lines show the module, or fiber tips plane, in different orientations}
        \label{fig:tangent_module}
	\end{subfigure}
 \begin{subfigure}[t]{0.33\textwidth}
		\centering
		\includegraphics[width=0.8\linewidth]{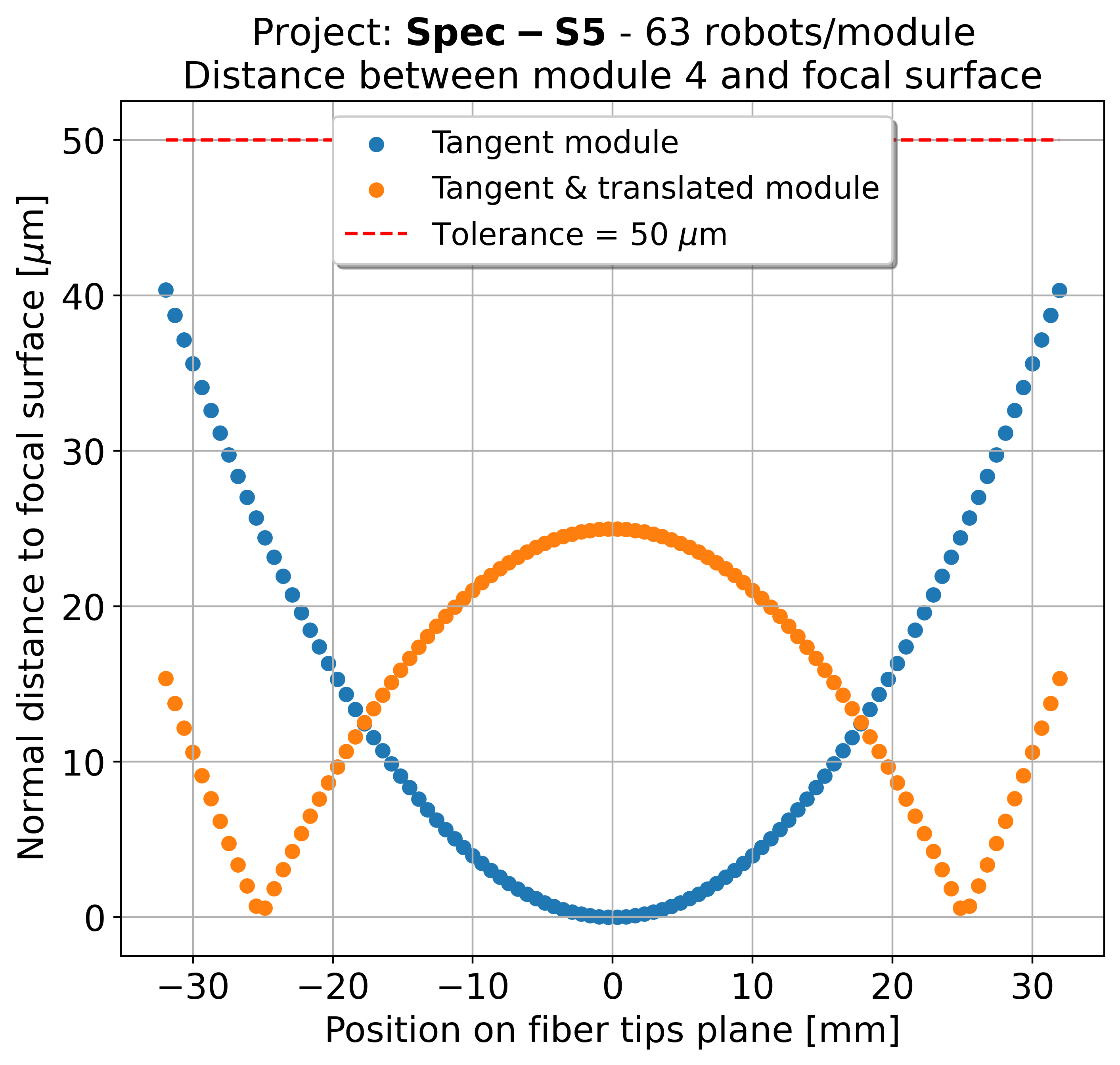}
		\caption{Normal distance for 63 robots}
        \label{fig:tangent_module_dist_63}
	\end{subfigure}
 	\begin{subfigure}[t]{0.33\textwidth}
		\centering
		\includegraphics[width=0.8\linewidth]{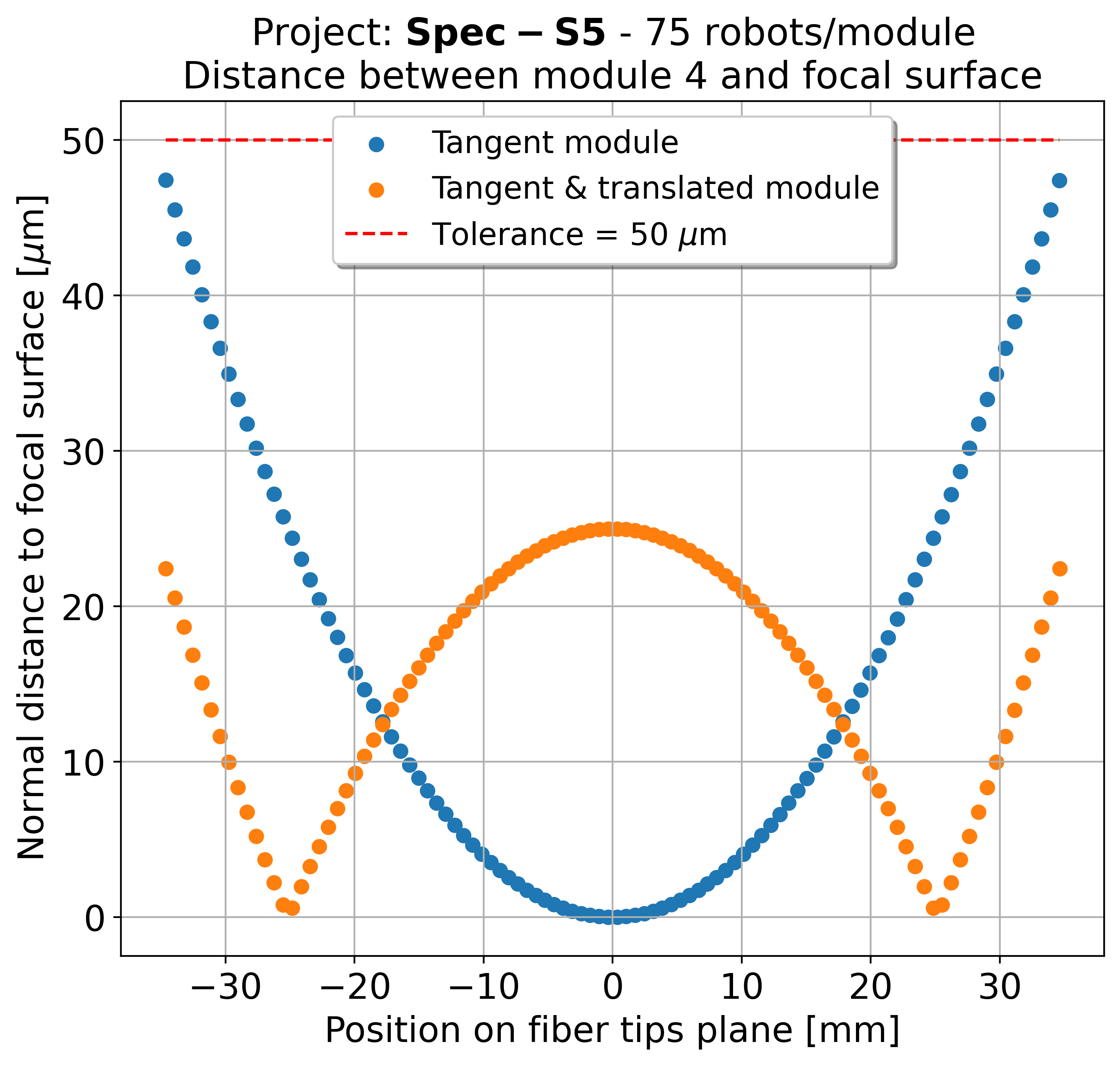}
		\caption{Normal distance for 75 robots}
        \label{fig:tangent_module_dist_75}
	\end{subfigure}
	\begin{subfigure}[t]{0.33\textwidth}
		\centering
		\includegraphics[width=0.8\linewidth]{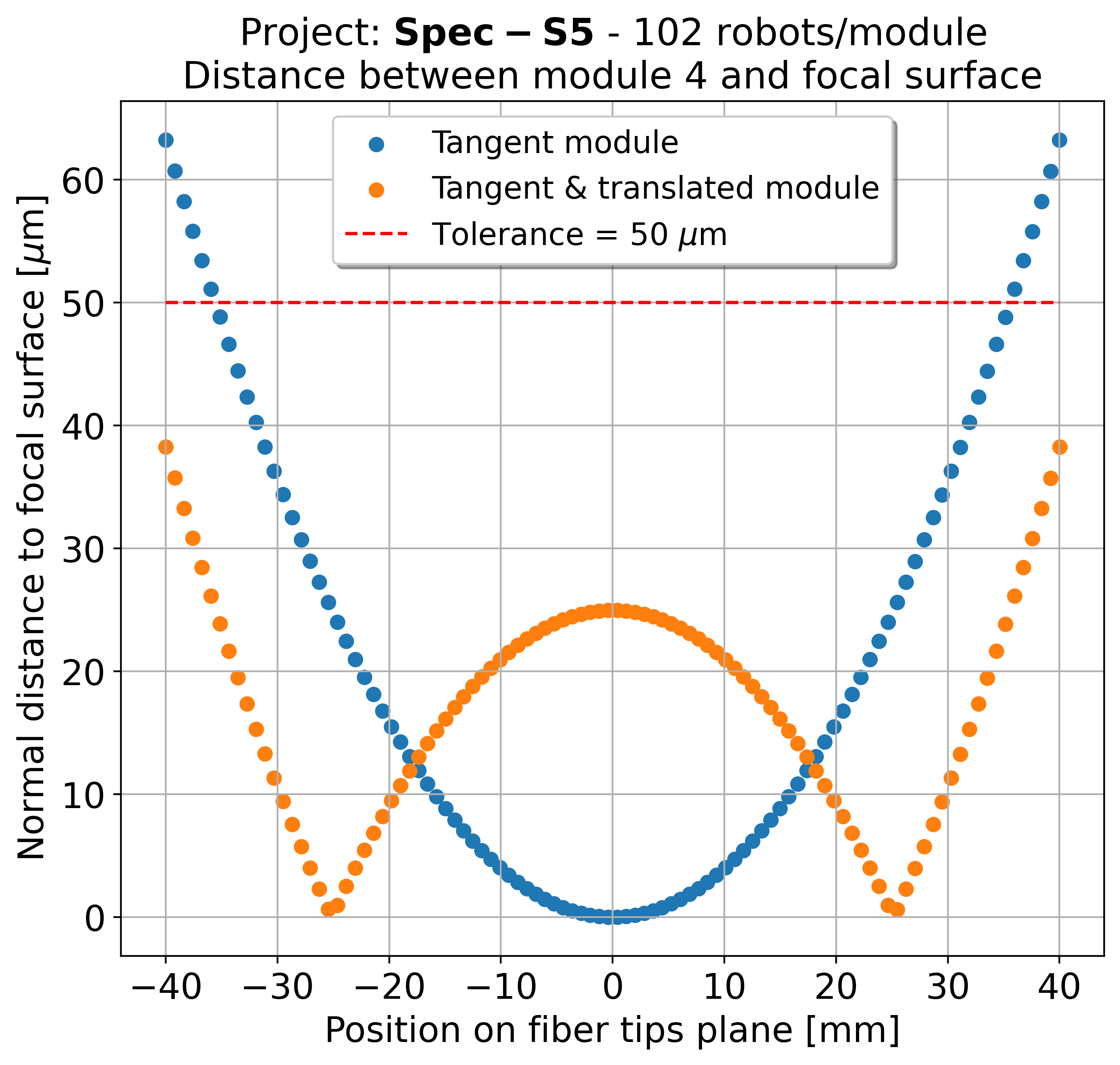}
		\caption{Normal distance for 102 robots}
        \label{fig:tangent_module_dist_102}
	\end{subfigure}
   \vspace{0.1cm}
	\caption{\ref{fig:tangent_module} shows of the fiber tips plane of a module placed at 222 mm from the center of the focal plane\\ Bottom figures show the normal distance of straight fiber tips plane for 63 (\ref{fig:tangent_module_dist_63}) and 102 robots (\ref{fig:tangent_module_dist_102})
 }
	\label{fig:fitting_comparison_nb_of_robs permodule} 
\end{figure}

First considerations on the number of robots per module can be done regardless of the three proposed options in Section \ref{sec:investigated_solutions} as this parameter impacts them in the same way. As stated in Section \ref{sec:nb_robots} it has a significant impact on the focal plane coverage but also on its capacity in fitting the focal surface, e.g. meeting the defocus and tilt tolerances for the fiber tips.\\
\\
As a first step, let us consider the fitting of the focal surface. Figure \ref{fig:tangent_module} illustrates the positioning of one module at 222 mm from its center. It is surrounded by a defocus tolerance envelope of $\pm$ 50 $\mu$m as a placeholder value since tolerance envelopes for defocus and tilt are not yet defined for proposed projects in Table \ref{tab:objects_projects}.Even as a placeholder this envelope illustrates that, in terms of focus only, a staired assembly depicted in Figure \ref{fig:staired_assembly} is not a suitable option for a modular assembly. Consequently, if the frameless concept were to be used the connection part would have to integrate the orientation of the module as well. Thus increasing even more the difficulty of meeting the necessary tolerances in module positioning that are already additioning as more module are assembled with a frameless concept.\\

\noindent Figure \ref{fig:tangent_module_dist_63} and \ref{fig:tangent_module_dist_63} represent the normal distance of the fiber tips plane across its length to the focal surface for the 63 and 102 robots cases. The said \textit{translated} ones were tangent then moved half of the tolerance envelope in their normal direction for them to lie inside the tolerance envelope.\\
They illustrate two ideas:
\begin{enumerate}
    \item 63 fiber tips can lie in the defocus tolerance envelope with a maximum margin of 10 $\mu$m by just being tangent to it, while 102 can not
    \item Translating the modules improves the position inside the defocus tolerance envelope allowing 25 $\mu$m of maximum margin while 102 only fits with a maximum margin of 10 $\mu$m
\end{enumerate}
They quantify the idea that increasing the number of robots per module indeed allows for more coverage but also jeopardizes the fitting of the focal surface. 

\noindent In parallel one could also calculate the angular difference between the light chief rays on the focal surface and the normal direction of the fiber tips plane similarly defined as in Figure \ref{fig:tangent_module} (for schematics in Figure \ref{fig:tilt_schematics}). Chief ray deviations being accounted for. 
\begin{table}[H]
\centering
\caption{Angular differences between normal direction of the fiber tips plane and the incoming chief rays across the length of said plane}
\label{tab:angular_diff_with_chief_ray}
\begin{tabular}{cccc}
\hline
           & Min ($\degree$) & Max ($\degree$) & Delta ($\degree$) \\ \hline
63 robots  & 0.24            & 0.67            & 0.43              \\
75 robots  & 0.26           & 0.74           & 0.48                  \\ \hline
102 robots & 0.31            & 0.87            & 0.56             
\end{tabular}
\end{table}

\begin{figure}[H]
\captionsetup[subfigure]{justification=centering}
	\begin{subfigure}[t]{0.5\textwidth}
		\centering
		\includegraphics[width=0.8\linewidth, trim=0 0 0 1.75cm, clip]{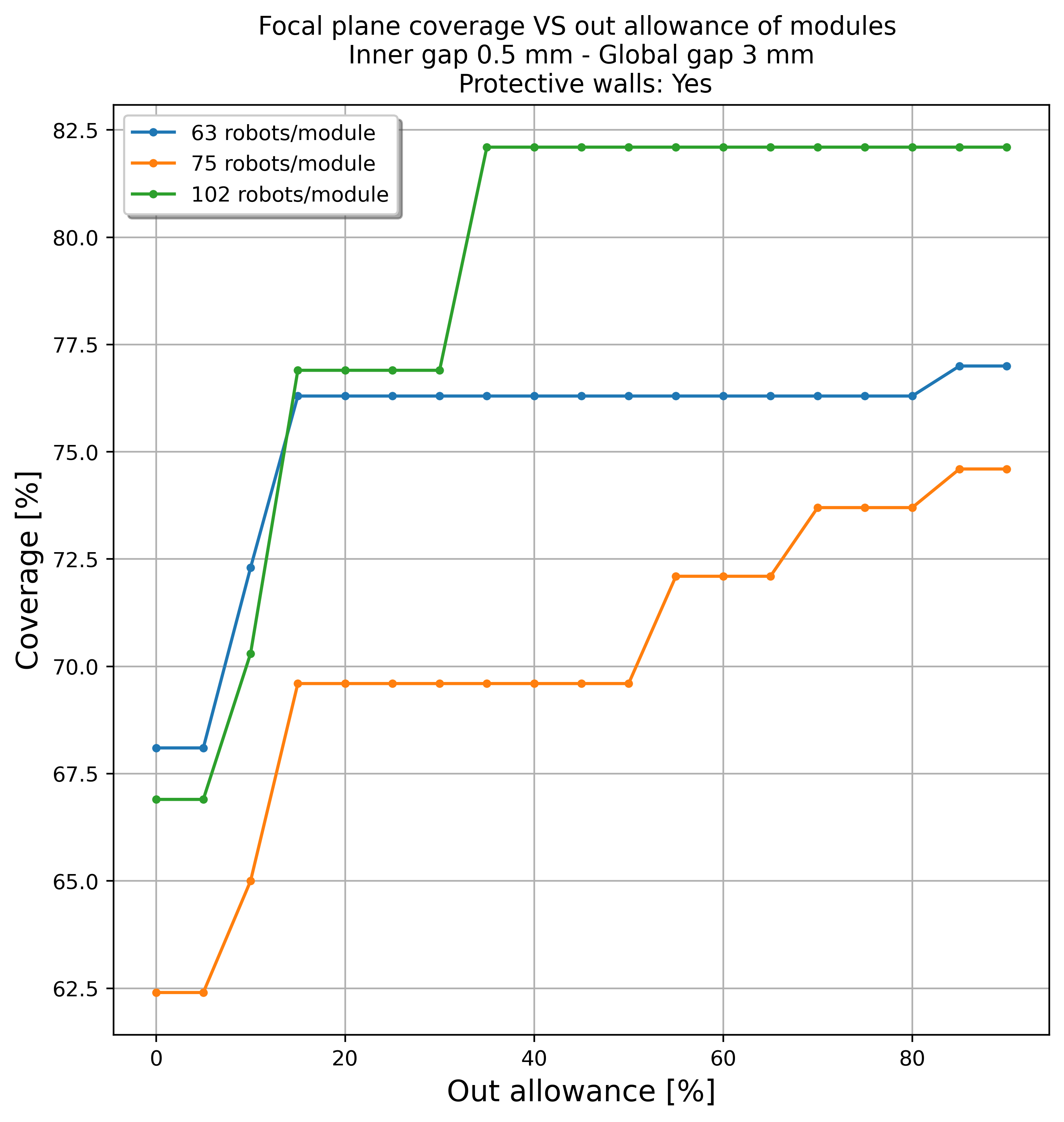}
		\caption{Semi-frameless: Inner gap: 0.5 mm - Global gap: 3 mm}
  \label{fig:coverage_plateau}
	\end{subfigure}
  \begin{subfigure}[t]{0.5\textwidth}
		\centering
		\includegraphics[width=0.8\linewidth, trim=0 0 0 1.75cm, clip]{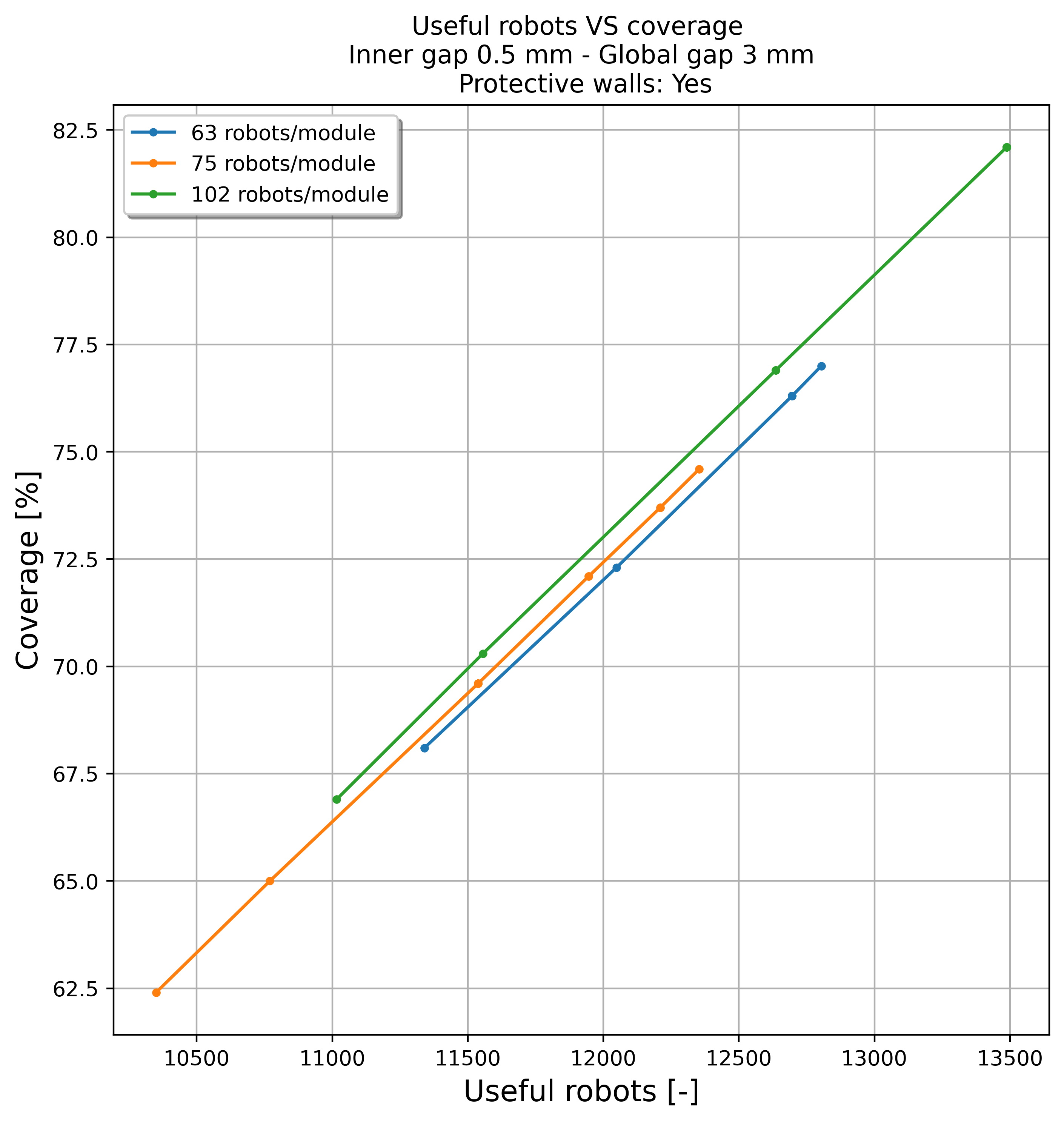}
		\caption{Coverage versus number of useful robots; each point represents a plateau from Figure \ref{fig:coverage_plateau}}
        \label{fig:coverage_useful}
	\end{subfigure}
  \hspace{1cm}
 \caption{Evolution of coverage with increasing \textit{out allowance} for a semi-frameless case (a) and number of useful robots achieving said coverage (b)}
 \label{fig:cov_comparison_concept}
\end{figure}

The second aspect of the choice on number of robots per modules concerns the coverage each case reach for a fixed assembly strategy, either framed, semi-frameless, frameless. Figure \ref{fig:cov_comparison_concept} highlights the evolution of coverage with respect to out allowance (a) and \textit{useful robots} (b), i.e. robots that contained within the vignetting radius, for a semi-frameless case.

The above study highlights first in Figure \ref{fig:coverage_plateau} the different plateau reached by coverage as out allowance increases. Maximum coverage is given by the 102 robots case with 82.4\% of the focal surface patrolled. The other two cases respectively 77.2\% and 73.2\% for 63 and 75 robots per module.\\
A first interpretation to such a difference is number of interfaces or walls created by the different options. Interfaces between module create as many blind zones as they are numerous. Bigger modules can easily cover the majority of the focal surface at once, explaining the performance of 102 case. An additional reason for the 75 to 63 difference are physical limits of module placements. In our case, they correspond to the GFAs onto which a module can not overlap, as stated in Section \ref{sec:pres_GFAs}. Since the smallest overlap removes an entire module, smaller modules have a better chance to fit within those limits. Thus, explaining the difference between 63 and 75 case for this particular set of parameters. Changing the module gaps can lead to a better layout of the 75 case, resulting in an inversion of this difference for coverage to follow properly the first explanation about the number of interfaces. Similar differences are found for the framed case.\\

One can find in Figure \ref{fig:coverage_useful} the linear relationship between coverage and the number of robots inside a vignetting radius, namely \textit{useful robots}. It highlights that for a similar number of useful robots 63 and 75 cases achieve similar coverage.\\

\noindent In conclusion, this section highlights three main points:
\begin{enumerate}
    \item 102 case while providing a much higher coverage than the other two cases, is not likely to be considered viable with respect to its low performances in fitting the focal surface
    \item 63 and 75 cases are similar in terms of performances for coverage and 63 case fits the focal surface the best
    \item Considering the trade-off coverage versus surface fitting as well as other factor such as ease of assembly, electronics or trillium fitting not covered in the scope of the study here: 63 robots per module is the optimal choice
\end{enumerate}

\subsection{Coverage comparison for the different concepts}
\begin{figure}[H]
\captionsetup[subfigure]{justification=centering}
\begin{subfigure}[t]{0.33\linewidth}
		\centering
		\includegraphics[width=\linewidth]{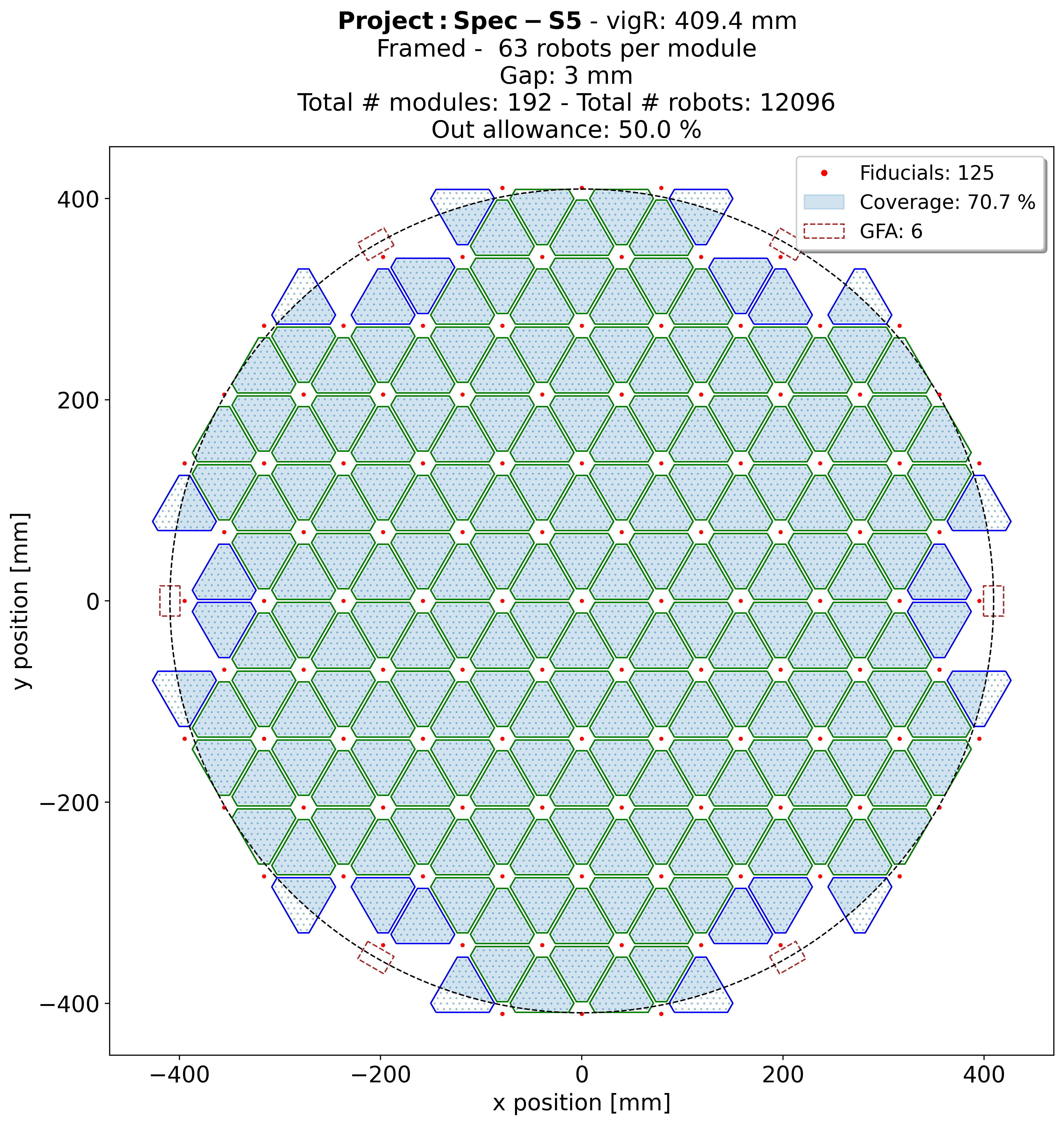}
		\caption{Framed layout\\
                Global gap: 3 mm - Coverage: 70.7 \%}
        \label{fig:cov_framed}
	\end{subfigure}
	\begin{subfigure}[t]{0.33\linewidth}
		\centering
		\includegraphics[width=\linewidth]{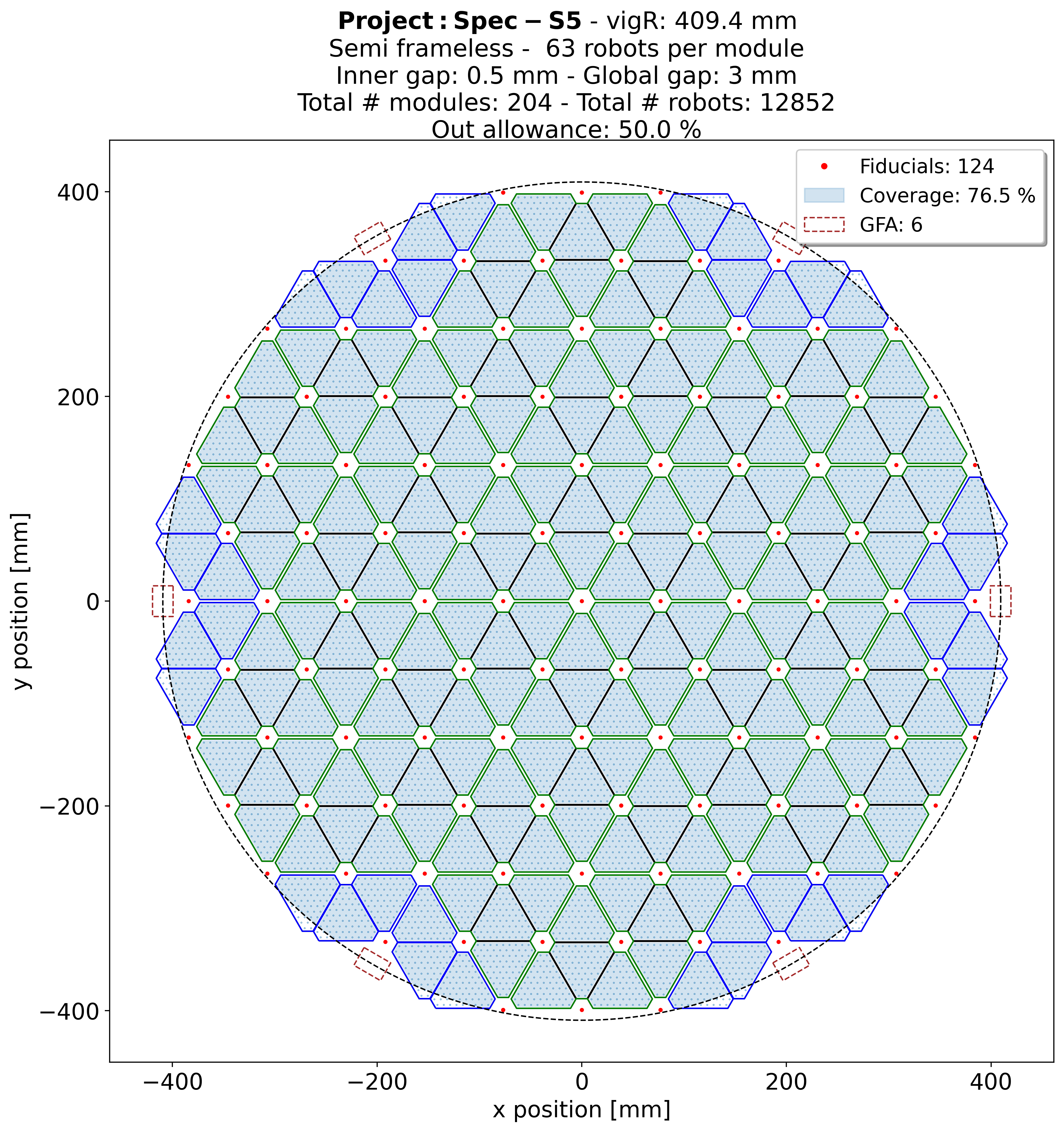}
		\caption{Semi-frameless layout\\
                Global gap: 3 mm - Inner gap: 0.5mm - Coverage: 76.3 \%}
        \label{fig:cov_semi_frameless}
	\end{subfigure}
 	\begin{subfigure}[t]{0.33\linewidth}
		\centering
		\includegraphics[width=\linewidth]{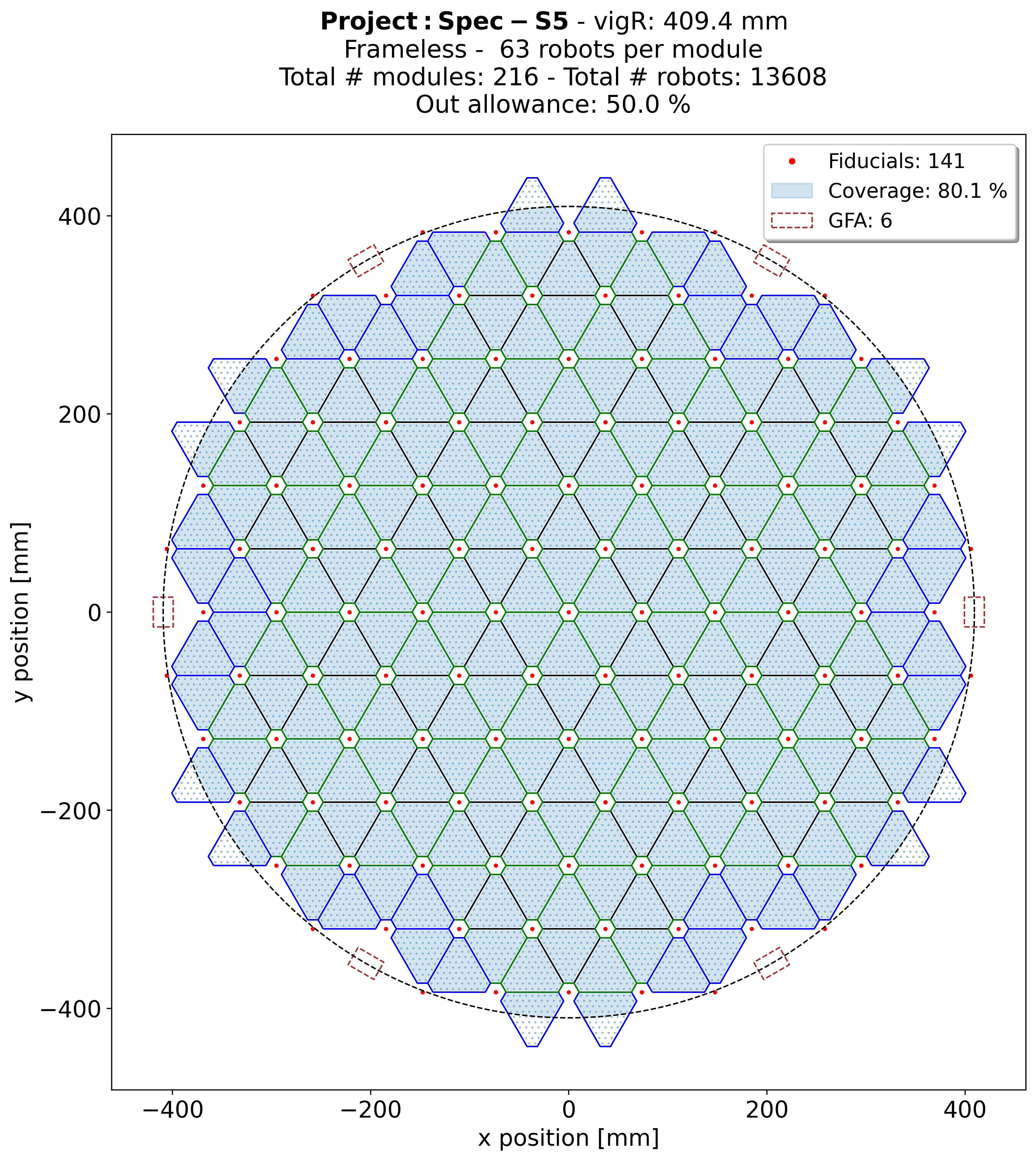}
		\caption{Frameless layout\\
                Global gap: 0 mm - Coverage: 80.1 \%}
        \label{fig:cov_frameless}
	\end{subfigure}
 \vspace{0.3cm}
	\caption{Coverage comparison between the framed, semi-frameless and frameless solutions - 63 robots case}
	\label{fig:coverage_framed_frameless}
\end{figure}

Figure \ref{fig:coverage_framed_frameless} illustrates the impact of module gap on a focal plane layout. The three represented cases are calculated for the same vignetting radius, out allowance parameters (50\%) and GFAs. The only changes occur in how close the modules are to each other. Figure \ref{fig:cov_framed} shows a similar assumption than for MM as for a 3 mm gap between modules resulting in a coverage of 70.7\% of the focal surface. Figure \ref{fig:cov_semi_frameless} implements the semi-frameless option keeping a global gap of 3 mm as previously but using an inner gap of 0.5 mm, as described in Section \ref{sec:semi-frameles solution}. Closing them together allows for 12 more modules to be added, thus increasing the coverage to 76.5\%. Finally, totally removing the gap between them in Figure \ref{fig:cov_frameless} provides space for 12 additional modules to fit within the out allowance parameter, rising the coverage to 80.1\%. The gap between the two extreme cases, (a) and (c) is therefore of 9.4\% and can be considered as a significant increase in coverage performance.\\

\noindent While those results are encouraging, one has to keep in mind two main limitations of the process:
\begin{enumerate}
    \item Coverage is but one aspect of the trade-off considered in this study, therefore even though the frameless concept might seem superior it does not seem to provide the necessary module positionning upon assembly
    \item Figure \ref{fig:coverage_framed_frameless} presents layouts made upon assumptions on certain parameters, especially the GFAs as discussed in Section \ref{sec:pres_GFAs}. Those results are then mostly to evolve as the projects gain in details on those aspects
\end{enumerate}
\newpage
\subsection{Finite Element Analyses for framed and semi-frameless concepts}
\label{sec:fea}
\begin{figure}[H]
\captionsetup[subfigure]{justification=centering}
\begin{subfigure}[t]{0.7\linewidth}
		\centering
		\includegraphics[width=\linewidth]{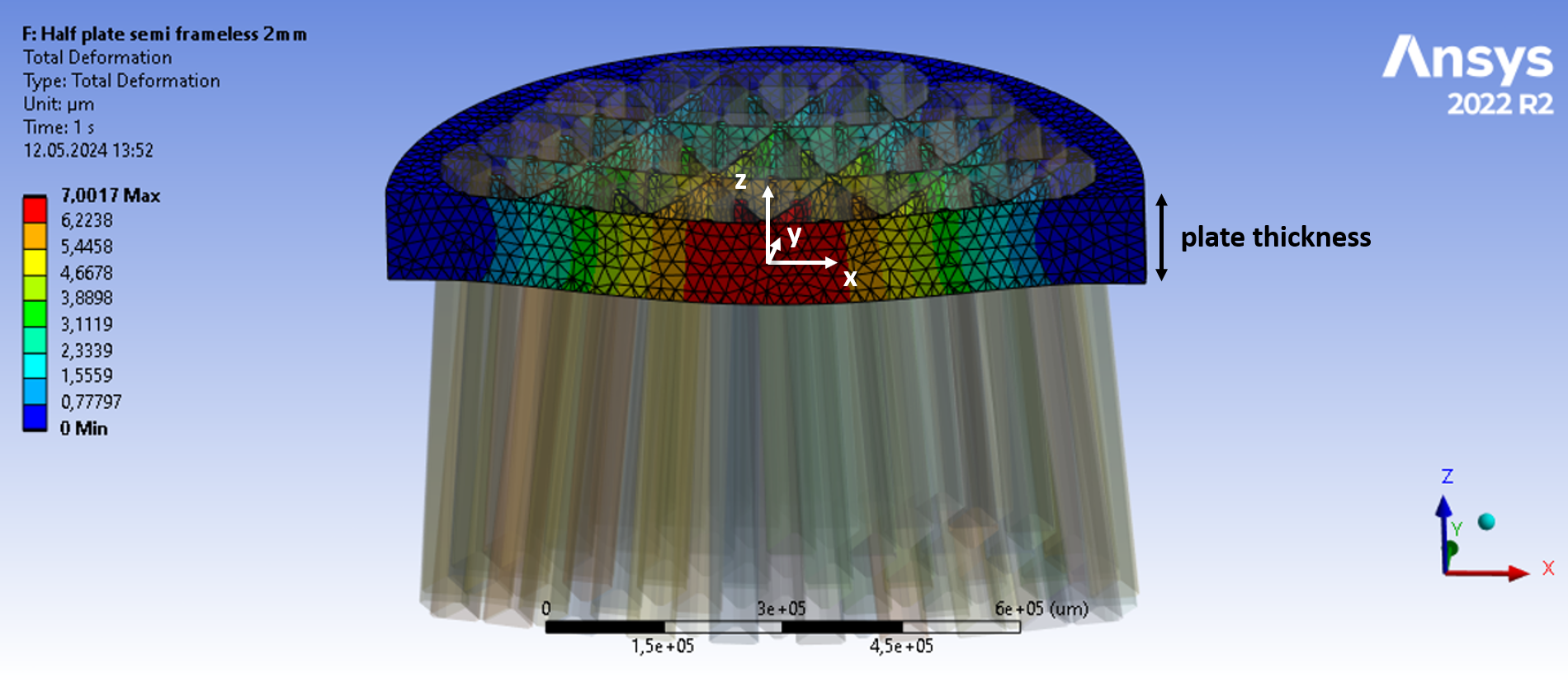}
		\caption{FEA of fully loaded 120 mm, stainless steel plate, looking at the zenith}
        \label{fig:FEA_global}
	\end{subfigure}
	\begin{subfigure}[t]{0.3\linewidth}
		\centering
		\includegraphics[width=0.7\linewidth]{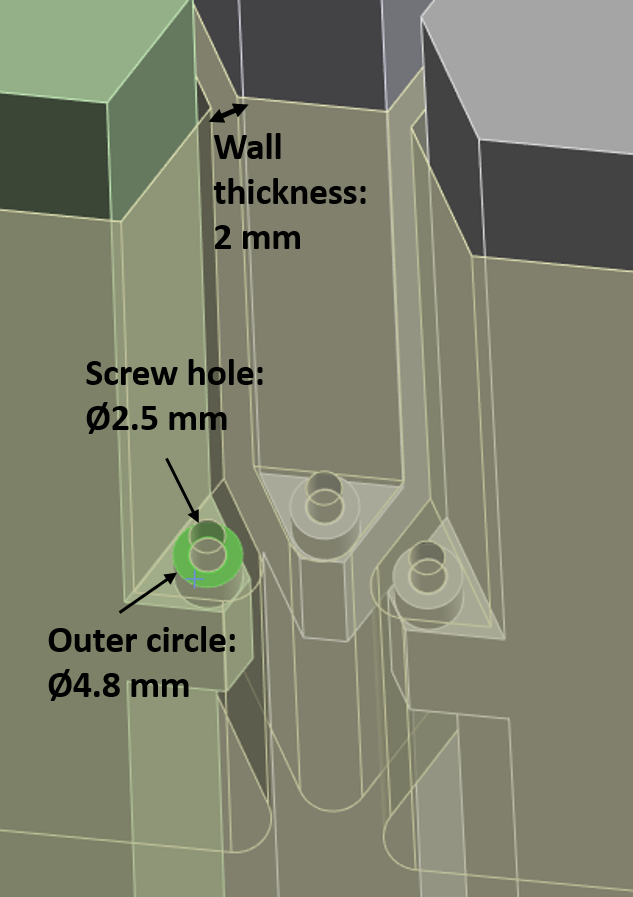}
		\caption{Zoomed view on contact surfaces modules/focal plate; green disk shows the bounded contact of one module}
	\end{subfigure}
  \hspace{1cm}
	\caption{FEA example for semi-frameless; 2 mm walls; 120 mm thick stainless steel plate; maximum deformation at the center: 7 $\mu$m}
	%\label{fig:inner_gap}
\end{figure}

In order to evaluate the performances of concepts with varying parameters Finite Elements Analyses (FEA) are realized on several focal plate models. A C\# routine was written by Xiangyu Xu to automate the generation of SolidWorks models and fasten the workflow of analyses which make the following assumptions:
\begin{itemize}
\item Boundary conditions:
\begin{itemize}
    \item Fixed outer surface of the plate
    \item Mid-plane symmetry instead of circular symmetry to easily account for the telescope rotation
    \item Bounded contact surface between modules and focal plate
\end{itemize}
\item Module:  the interface linking the modules and the focal plate is assumed to be SStl. A simplified geometry of module is used for calculation purposes. A custom material is used to simulate both SStl stiffness and maximum target mass (1.6 kg) with a custom density for the right mass distribution. The characteristics used for the materials are summarized in Table \ref{tab:materials_fea}
\item Only gravity load is considered
\item Rotation of the whole assembly occurs about the principal y axis from the coordinate system placed at the center of the assembly, see Figure \ref{fig:FEA_global}

\end{itemize}

% \usepackage{booktabs}
\begin{table}[H]
\centering
\caption{Material properties used for FEAs}
\label{tab:materials_fea}
\begin{tabular}{@{}cccc@{}}
\toprule
                       & Young's Modulus (GPa) & Poisson's ratio (-) & Density (kg/m$^3$) \\ \midrule
Stainless Steel (SStl) & 190                   & 0.31                & 7750               \\
Aluminium (Alu)        & 71                    & 0.33                & 2700               \\
Custom module material & 190                   & 0.31                & 5200               \\ \bottomrule
\end{tabular}
\end{table}

Figure \ref{fig:FEA_global} shows the basis of the FEAs performed on the fully loaded plated. The automated plate model generation allowed for easy repetition of similar load cases for both semi-frameless and framed concepts on varying \q{walls thicknesses} i.e. the thickness of each \q{web string}.\\
% \begin{figure}[H]
% \captionsetup[subfigure]{justification=centering}
% \begin{subfigure}[t]{0.5\textwidth}
% 		\centering
% 		\includegraphics[width=0.8\linewidth]{figures/004_evaluating_solutions/max_def_alu.png}
% 		\caption{Aluminium plate;}
%         \label{fig:max_def_alu}
% 	\end{subfigure}
% 	\begin{subfigure}[t]{0.5\textwidth}
% 		\centering
% 		\includegraphics[width=0.8\linewidth]{figures/004_evaluating_solutions/max_def_stainless.png}
%         \caption{Stainless steel plate}
%         \label{fig:max_def_stainless}
% 	\end{subfigure}
%  \caption{Maximum focal plate total deformation when looking at the zenith for aluminium (a) and stainless steel plate (b) for varying plate thicknesses}
%  \end{figure}
\subsubsection{Plate deformation}
\begin{figure}[H]
\captionsetup[subfigure]{justification=centering}
\begin{subfigure}[t]{0.47\linewidth}
    \centering
    \includegraphics[width=\linewidth]{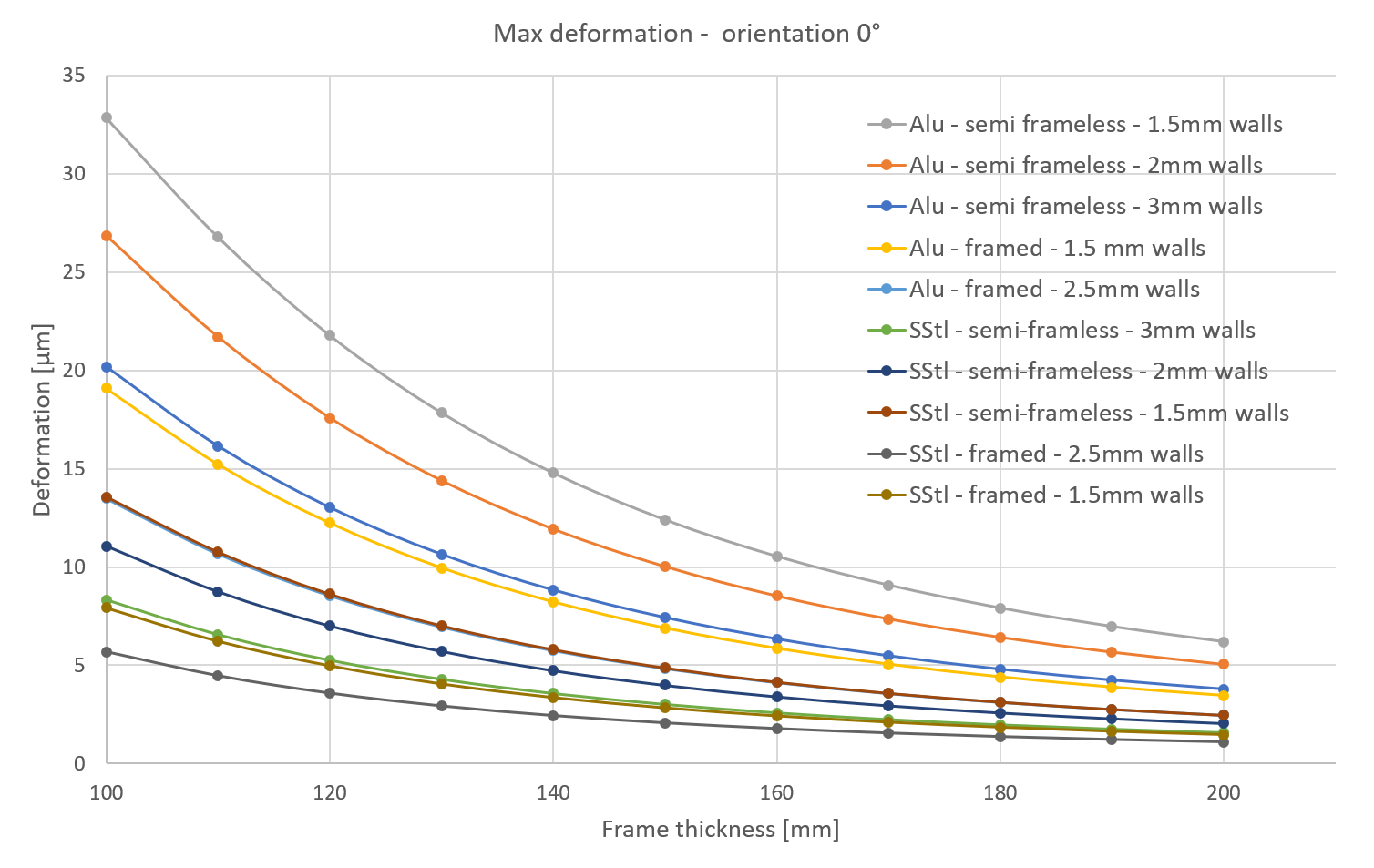}
    \caption{Maximum deformation of fully loaded plates at the zenith for varying thicknesses}
    \label{fig:max_def_thickness}
	\end{subfigure}
 \begin{subfigure}[t]{0.53\linewidth}
    \centering
    \includegraphics[width=\linewidth]{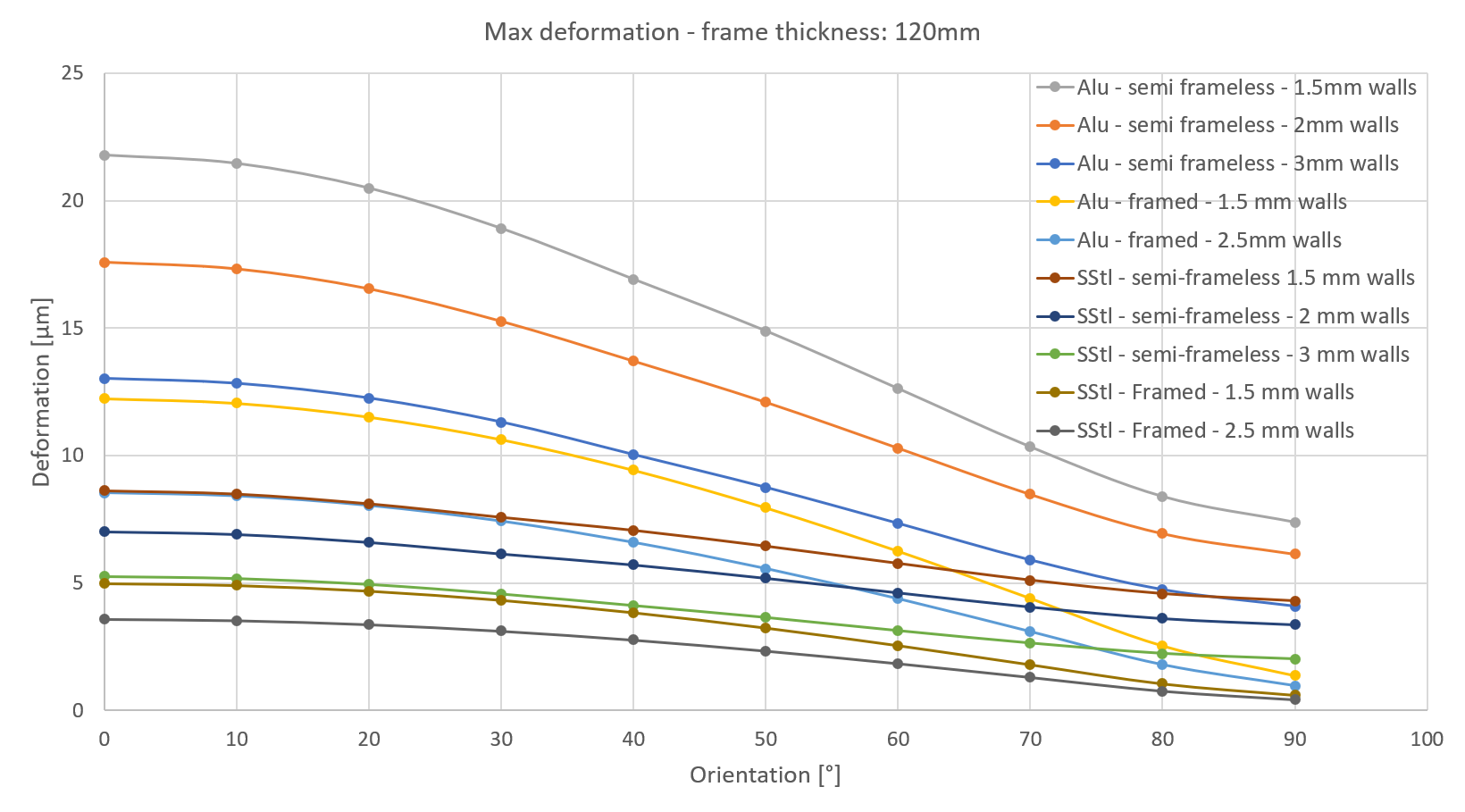}
    \caption{Maximum deformation of 120 mm thick plates over a range of 0 to 90 \textdegree}
    \label{fig:max_def_oriented}
    \end{subfigure}
    \vspace{0.3cm}
    \caption{Maximum expected deformation of the focal plates for varying thicknesses and fixed orientation at zenith (a) and for varying orientations and a fixed thickness of 120 mm (b); "Alu" and "SStl" respectively stand for "Aluminium" and "Stainless Steel"}
\end{figure}

 \noindent This first analyses batch in Figure \ref{fig:max_def_thickness} shows the expected maximum deformations of loaded plates versus the thickness of the plate to assess the correct dimension needed depending on the project requirement. The study is on a voluntarily broad range to give a global picture of the performances of the different concepts. It can then be refined once the requirements are defined.\\
 The first noticed behavior is the convergence of all the cases as the plates thicken. The stiffness of the assembly is dominated by the structure itself at first, the thickest framed plate deforming the less. As the plate thickens the material itself dominates the plate stiffness more than its structure and all the cases for a same material converge.\\
 The change of material has an impact on the final results. For example, two 100 mm thick semi-frameless plate of 1.5mm walls the maximum deformation for aluminium is 33 $\mu$m and 14 $\mu$m for the stainless steel. Therefore, a stainless steel plate deforms $\approx$ 2.3 times less than the aluminium when looking at the zenith.\\

 \noindent For the considered projects, the telescopes need to rotate their focal plane during the observation. It is therefore important to assess their deformation during such motion to ensure that it remains within the requirements. Such a study can be found in Figure \ref{fig:max_def_oriented}.\\
 It shows the evolution of the maximum plate deformation as the assembly rotates with the telescope and states that, as a worst case for an aluminium semi-frameless plate with walls of 1.5mm, the expected deformation is bounded between 22 $\mu m$ and 7.5 $\mu m$.\\

% \begin{figure}[H]
% \captionsetup[subfigure]{justification=centering}
%  \begin{subfigure}[t]{0.5\textwidth}
% 		\centering
% 		\includegraphics[width=0.8\linewidth]{figures/004_evaluating_solutions/max_def_alu_angle.png}
% 		\caption{Aluminium plate}
% 	\end{subfigure}
% 	\begin{subfigure}[t]{0.5\textwidth}
% 		\centering
% 		\includegraphics[width=0.9\linewidth]{figures/004_evaluating_solutions/max_def_stainless_angle.png}
% 		\caption{Stainless steel plate}
% 	\end{subfigure}
%   \hspace{1cm}
%  \caption{Maximum focal plate total deformation when assembly is rotating for a fixed plate thickness of 120 mm and varying material}
%  \label{fig:max_def_angle}
% \end{figure}

\subsubsection{Modules tilt}
\label{sec:tilts}
\begin{figure}[H]
\captionsetup[subfigure]{justification=centering}
 \begin{subfigure}[t]{0.4\textwidth}
		\centering
		\includegraphics[width=0.9\linewidth]{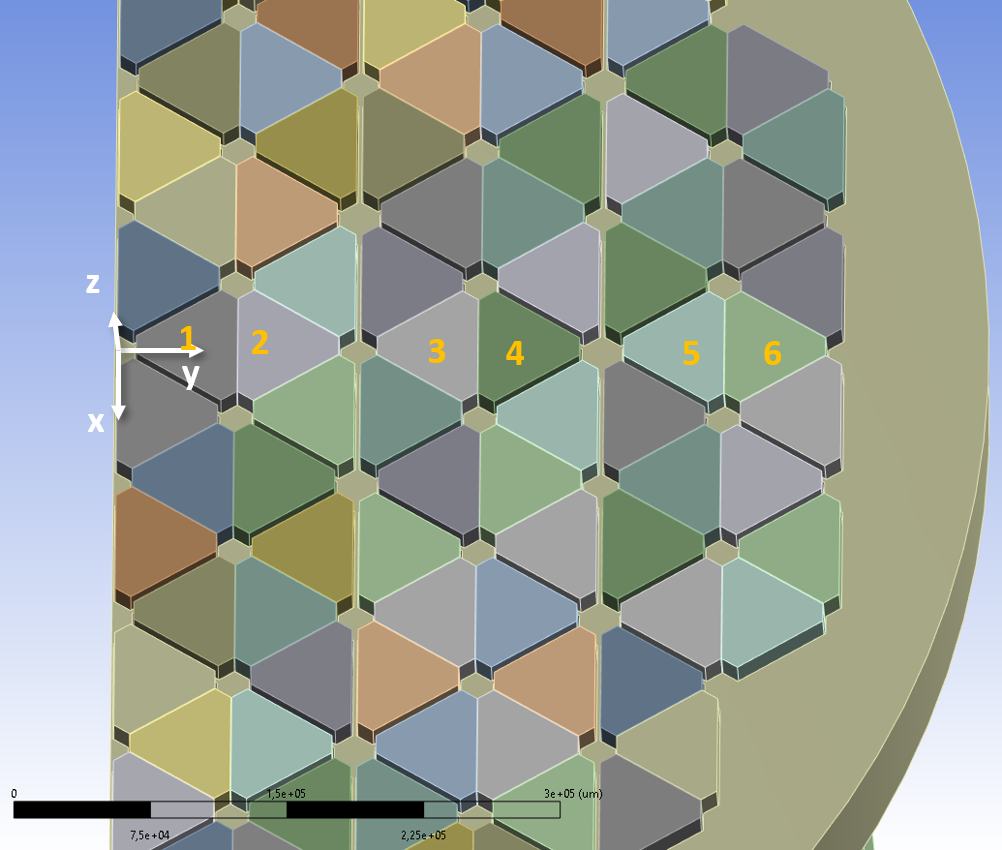}
		\caption{Modules considered in the tilt FEAs; tilting is considered about the x axis only when looking at the zenith}
	\end{subfigure}
	\begin{subfigure}[t]{0.6\textwidth}
		\centering
		\includegraphics[width=\linewidth]{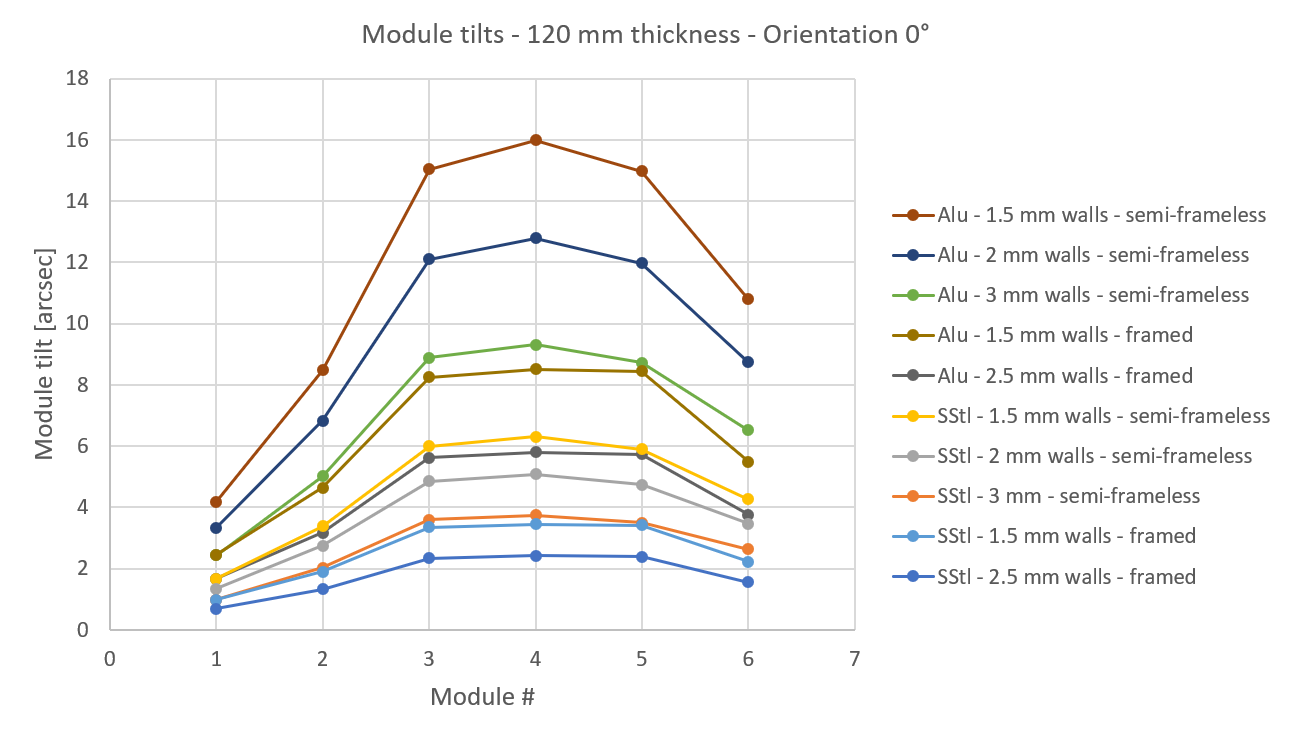}
        \caption{Tilt angles of the considered modules once the plates are fully loaded and oriented to the zenith}
        \label{fig:max_tilt}
	\end{subfigure}
  \vspace{0.1cm}
 \caption{Expected tilt angles of a few modules across the radius of the focal plates}
 \label{fig:max_def_angle}
\end{figure}

This tilt study of the modules highlights the same trend as the plate deformation. It is important to note that even though the maximum deformation occurs at the center of the plate in Figure \ref{fig:FEA_global}, the $\#1$ module is not the one that tilts the most. The highest tilt can be found for the $\#3$, $\#4$ and $\#5$ modules, reaching up 16'' $\approx$ 0.0044$\degree$ for the less stiff case. As no tolerances are defined for future projects yet, one can first compare those tilts with the nominal angular deviation reported in Table \ref{tab:angular_diff_with_chief_ray} and note that tilts are two order of magnitude smaller. Therefore, this study highlights the minimal impact that the tilts induced by the deformation of the plate will have in the angular tolerance chain.

 Those studies show that, expectedly, the framed case is stiffer than the semi-frameless one. By giving the global overview, they allows to quantify this difference. Based on the requirements of each project, e.g. how much coverage is wanted versus the allowed defocus tolerance, one can now choose which concept fits them the best.

\subsection{Prototyping of small semi-frameless assembly}
\begin{figure}[H]
\captionsetup[subfigure]{justification=centering}
 \begin{subfigure}[t]{0.5\textwidth}
		\centering
		\includegraphics[width=0.7\linewidth]{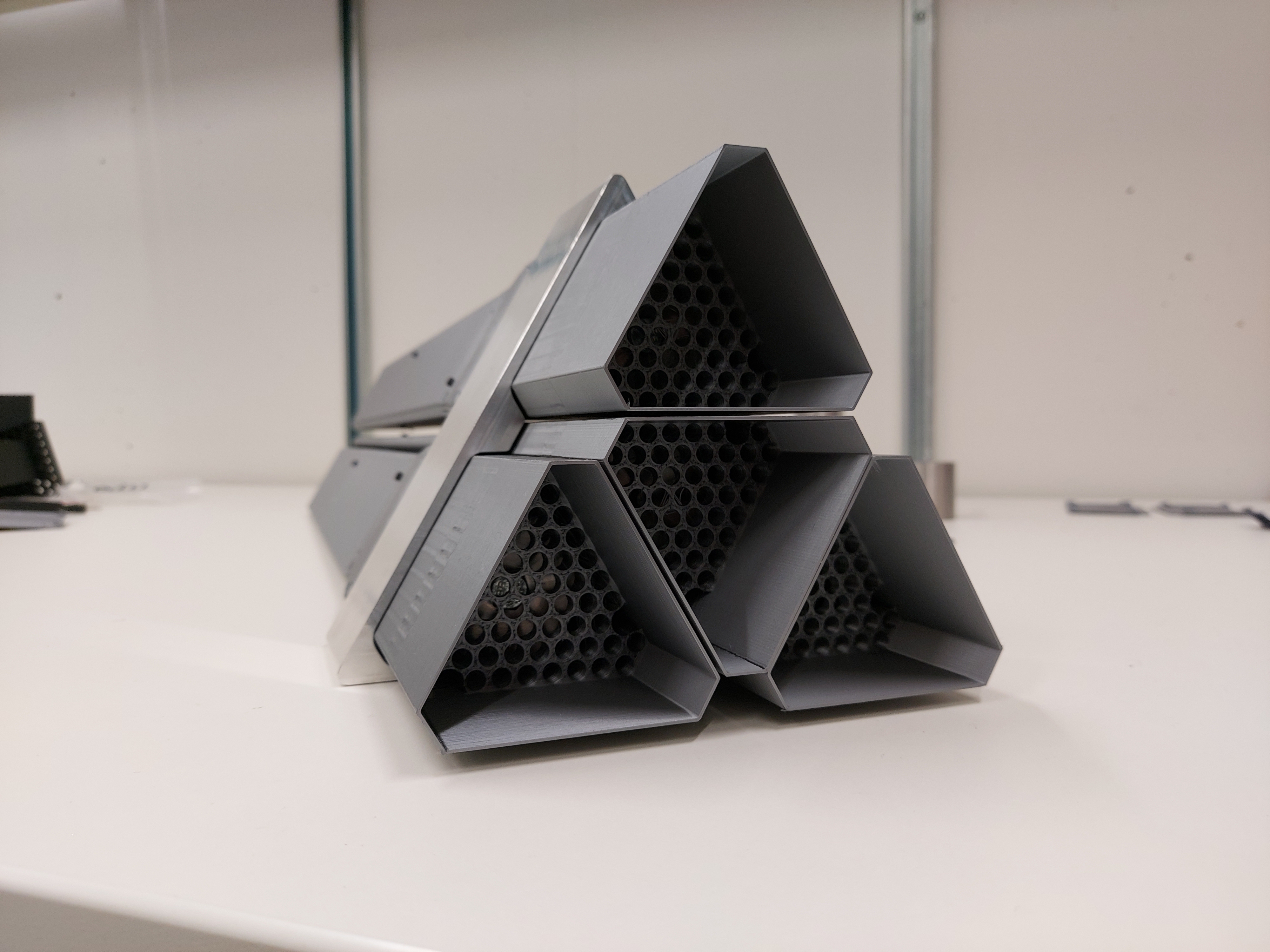}
		\caption{Front view of the assembled modules, robots side}
        \label{fig:proto_front_view}
	\end{subfigure}
	\begin{subfigure}[t]{0.5\textwidth}
		\centering
		\includegraphics[width=0.7\linewidth]{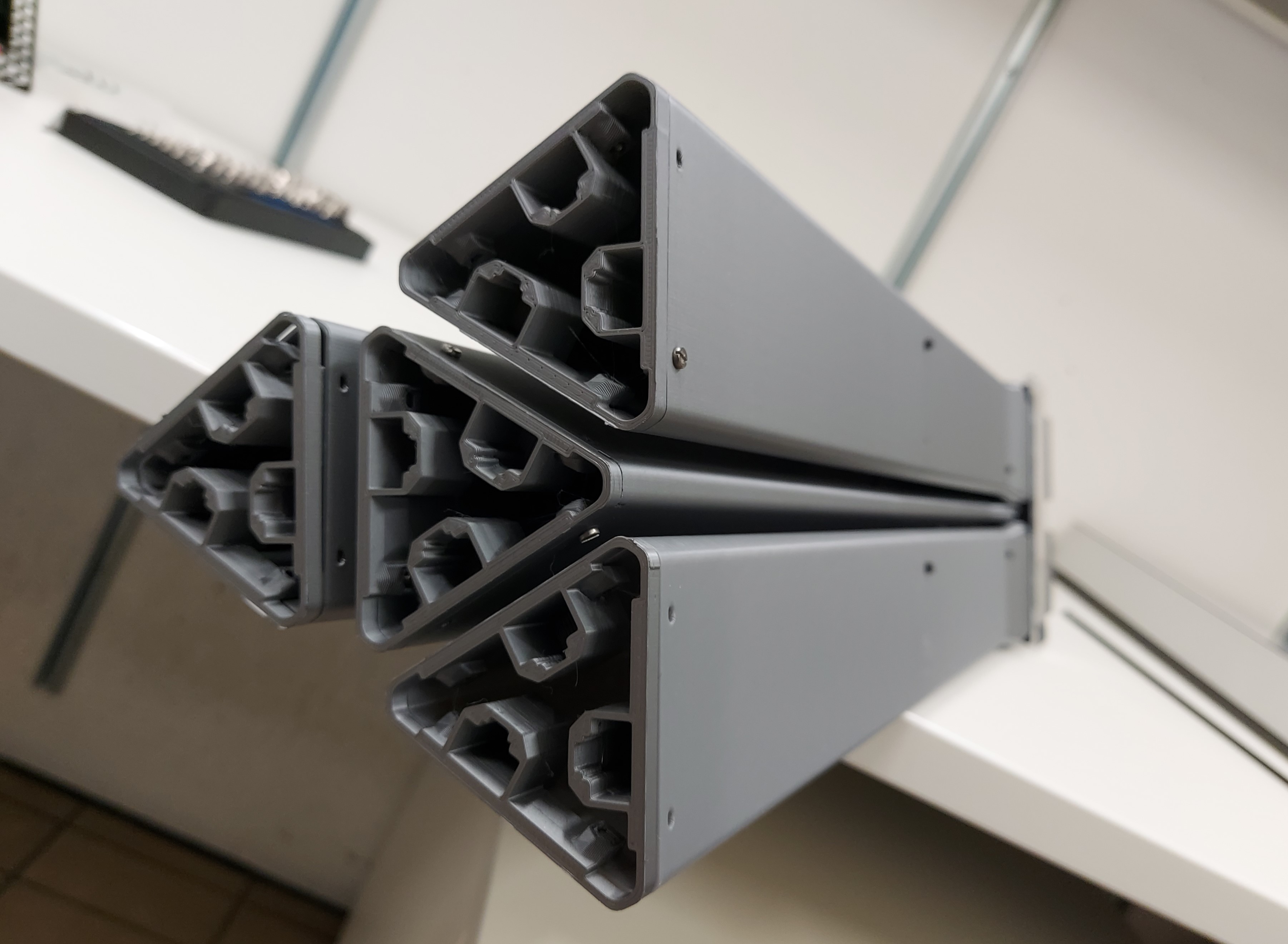}
        \caption{Back view, connectors and fibers routing side}
        \label{fig:proto_back_view}
	\end{subfigure}
 	\begin{subfigure}[t]{\textwidth}
		\centering
		\includegraphics[width=0.3\linewidth]{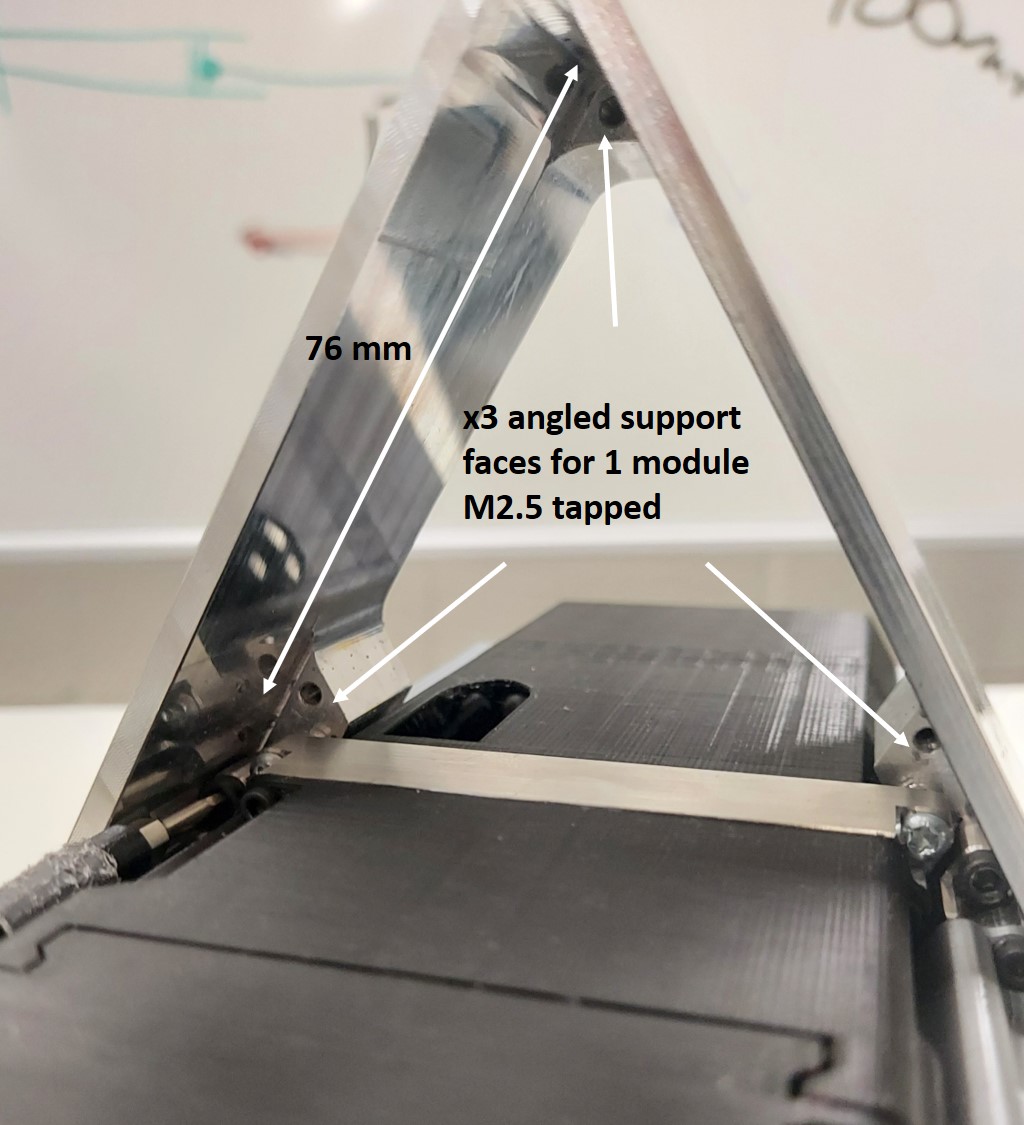}
        \caption{Inside view of the prototyped frame}
        \label{fig:proto_inside_view}
	\end{subfigure}
  \vspace{0.1cm}
 \caption{Prototype of semi-frameless assembly; dummy modules are 600 mm long}
 \label{fig:prototype_assembly}
\end{figure}

An aluminium semi-frameless prototype and dummy modules were manufactured to test the ease of assembly and milling precision. They are bolted to the frame using a custom made module-long screw driver and three M2.5 screws similarly to what was proposed in MM\cite{blanc_megamapper_2022}. The support faces were manufactured to give the three outer modules a nominal angle of 0.23$\degree$ with respect to central one, manufactured flat. Visual confirmation of this angle can be noticed between Figure \ref{fig:proto_front_view} and \ref{fig:proto_back_view} were the front side of the modules are closer than the back side. Measurements performed with a Coordinate Measurement Machine allowed to assess the angular RMS error of the support faces at 0.08$\degree$. This value is larger than the ones discussed in Section \ref{sec:tilts} which shows that improvements are needed in the manufacturing process and the usefulness of the shims stack envisioned at each edge of the modules to allows for tilt corrections upon assembly.

%% file: 005_conclusion.tex
\newpage
\section{Conclusion}
This study allowed to draw conclusions for future designs of modules of fiber positioners the focal plate structure that is holding them in place in the instrument. First, the coverage versus focal surface fitting trade-off considerations lead towards an optimum of 63 robots per module. The reduced size compared to the 75 robots MM baseline allows to fit the surface better while maintaining a good coverage. On their assembly in the instrument, the frameless concept does not seem to optimal in terms of positioning tolerances. As the modules are linked together their tolerance chain is going to grow to fast most likely out of specs. Therefore, leading to the investigation on an intermediate solution of the semi-frameless assembly. Locally bringing closer modules increases the coverage while maintaining good performances of deflections and tilt. A stiffness decrease is still to be noticed compared to the framed case. This overview gives one the global picture of the different performances to help choosing between the concepts and dimensions once specifications are set for the future multiplexed instruments.